\pdfoutput=1
\documentclass{article}

\usepackage{iclr2027_conference, times}
\iclrfinalcopy

\usepackage[utf8]{inputenc} %
\usepackage[T1]{fontenc}    %
\usepackage{url}
\usepackage{booktabs}       %
\usepackage{amsfonts}
\usepackage{nicefrac}       %
\usepackage{microtype}
\usepackage{xcolor} 
\usepackage[font=small,labelfont=bf]{caption}
\usepackage[inline]{enumitem}
\usepackage{textcomp}
\usepackage{makecell}
\usepackage{fontawesome5}
\usepackage{placeins}
\usepackage{inconsolata}

\usepackage{amsmath, amssymb}
\usepackage{graphicx}
\usepackage{wrapfig} 
\usepackage{siunitx}
\newcommand{\best}[1]{\bfseries\num{#1}}
\newcommand{\second}[1]{\multicolumn{1}{c}{\underline{\num{#1}}}}

\usepackage{tikz}

\usepackage{subcaption}

\usepackage{amsthm}
\newtheorem{theorem}{Theorem}

\newtheorem{proposition}[theorem]{Proposition}

\newtheorem{definition}{Definition}

\usepackage[final]{changes}

\definechangesauthor[name={Florian}, color=tabolive]{fg}
\definechangesauthor[name={Roland}, color=tabblue]{rk}
\definechangesauthor[name={Dominik}, color=tabcyan]{dg}
\setdeletedmarkup{\sout{\textcolor{red}{#1}}}

\usepackage{hyperref}       %
\hypersetup{
    colorlinks=true,
    linkcolor=modernteal,
    citecolor=modernteal,
    urlcolor=modernteal
}

\usepackage[capitalise]{cleveref}

\crefname{equation}{Eq.}{Eqs.}
\Crefname{equation}{Eq.}{Eqs.}
\crefname{definition}{Definition}{Definitions}
\Crefname{definition}{Definition}{Definitions}
\crefname{appendix}{Appendix}{Appendices}
\Crefname{appendix}{Appendix}{Appendices}

\usepackage{doi}
\renewcommand{\doi}[1]{%
  \href{https://doi.org/#1}{\texttt{doi:#1}}%
}

\DeclareMathOperator{\dgm}{dgm}

\DeclareMathOperator{\Del}{Del}
\DeclareMathOperator{\Flood}{Flood}
\DeclareMathOperator{\mFlood}{mFlood}
\DeclareMathOperator{\hull}{conv}

\DeclareRobustCommand{\rg}{\texttt{RG}}
\DeclareRobustCommand{\ssf}{\texttt{SSF}}
\DeclareRobustCommand{\fnc}{\texttt{FNC}}
\DeclareRobustCommand{\rmsd}{\texttt{RMSD}}

\newcommand{\floodph}{\texttt{Flood PH}}
\newcommand{\mfloodph}{\texttt{mFlood PH}}
\newcommand{\ripsphca}{\texttt{Rips PH} (C$_\alpha$)}
\newcommand{\alphaphca}{\texttt{Alpha PH} (C$_\alpha$)}
\newcommand{\alphaph}{\texttt{Alpha PH}}
\newcommand{\marsfm}{\texttt{MarS-FM}}
\newcommand{\marsfmmflood}{\texttt{MarS-FM-mFlood}}
\newcommand{\marsfmflood}{\texttt{MarS-FM-Flood}}
\newcommand{\marsfmalphaca}{\texttt{MarS-FM-Alpha} (C$_\alpha$)}
\newcommand{\marsfmalpha}{\texttt{MarS-FM-Alpha}}
\newcommand{\mdcath}{\texttt{mdCATH}}

\newcommand{\std}[1]{\textcolor{gray}{\scriptsize\textpm\ #1}}
\definecolor{tabblue}{RGB}{31,119,180}
\definecolor{taborange}{RGB}{255,127,14}
\definecolor{tabgreen}{RGB}{44,160,44}
\definecolor{tabred}{RGB}{214,39,40}
\definecolor{tabpurple}{RGB}{148,103,189}
\definecolor{tabbrown}{RGB}{140,86,75}
\definecolor{tabpink}{RGB}{227,119,194}
\definecolor{tabgray}{RGB}{127,127,127}
\definecolor{tabolive}{RGB}{188,189,34}
\definecolor{tabcyan}{RGB}{23,190,207}
\definecolor{introred}{rgb}{1,0,0.023}
\definecolor{introgray}{rgb}{0.901, 0.901, 0.956}
\definecolor{fig3blue}{rgb}{0.21, 0.464, 0.684}

\definecolor{citationblue}{RGB}{0,80,160}
\definecolor{linkblue}{HTML}{1A1AFF}   %
\definecolor{citegreen}{HTML}{117733}  %
\definecolor{urlred}{HTML}{AA0000}     %
\definecolor{verylightyellow}{rgb}{1.0, 1.0, 0.88}
\definecolor{modernred}{HTML}{bc272d}
\definecolor{modernteal}{HTML}{0000a2}
\definecolor{modernpink}{HTML}{E91E63}

\title{Learning Topological Representations \\of Protein Structure and  Dynamics}

\author{Dominik Geng$^{1}$, Florian Graf$^{1}$, Martin Uray$^{1,2}$, Roland Kwitt$^{1}$\\
{\small\mdseries $^{1}$University of Salzburg, Austria}\\
{\small\mdseries $^{2}$Josef Ressel Centre for Intelligent and Secure Industrial Automation,}\\
{\small\mdseries\phantom{$^{2}$}University of Applied Sciences, Salzburg, Austria}}

\makeatletter
\newsavebox\anontitlebox
\def\@maketitle{%
  \sbox\anontitlebox{\vbox{\hsize\textwidth
    {\LARGE\sc \@title\par}%
    \begin{tabular}[t]{l}\bf\rule{\z@}{24pt}Anonymous authors\\Paper under double-blind review\end{tabular}%
    \vskip 0.3in}}%
  \vbox to \ht\anontitlebox{\hsize\textwidth
    \vskip -14pt%
    {\LARGE\sc \@title\par}%
    \begin{tabular}[t]{l}\bf\rule{\z@}{24pt}\@author\end{tabular}%
    \vfil}}
\makeatother

\begin{document}

\maketitle
\lhead{Preprint}
\begin{abstract}

Modern protein representation models support tasks such as enzyme design and drug discovery, but their reliance on static data such as sequence and native structure limits their ability to capture the conformational dynamics that drive protein function. We investigate whether persistent homology (PH) can provide descriptors shared across diverse proteins that retain global structure, fine-grained conformational variability, and kinetically relevant information without large-scale pretraining. We introduce the \emph{masked Flood complex}, i.e., an adaptation of a recently proposed simplicial complex construction, that incorporates domain knowledge to emphasize inter-residue structure at low computational cost. We then use it to compute PH on molecular dynamics (MD) sampled structures, vectorize the persistence diagrams into a shared coordinate system, and probe the capacity of these representations in terms of the aforementioned aspects. To assess the amount of kinetic information, we learn low-dimensional embeddings from time-lagged observations and evaluate Markov state models (MSMs) estimated from them. Using these MSMs to guide training of the recent \marsfm\ generative framework improves several ensemble statistics relative to the original model. After finetuning on lower-temperature MD data and adapting the sampling procedure, the resulting model also shows promising transfer to fast folding proteins.

\end{abstract}

\section{Introduction}
\label{section:introduction}

Machine learning has driven major advances in protein
structure prediction \citep{Jumper.2021,Aranganathan.2025}.
Complementing these advances, representations learned from sequence
\citep{Lin.2023}, structure \citep{Hsu.2022}, and their integration
with functional annotations \citep{Hayes.2025} support the study of
protein function, with applications to screening
candidate drugs, antibodies, and enzymes
\citep{Nguyen.2025,Heinzinger.2025}.

While protein structure is a primary determinant of function, it is
inherently \emph{dynamic}: even the native state of a protein
comprises an ensemble of rapidly fluctuating conformations with
comparable stability \citep{Voet.2016,McCammon.1977}, while
intrinsically disordered proteins explore heterogeneous
ensembles \citep{Trivedi.2022}.
These dynamics underlie processes such as protein folding
\citep{Jaeger.2006}, transient cryptic-pocket formation
\citep{Zhang.2026}, and ligand binding \citep{Ostermann.2000}.

A standard computational approach to study these processes is
\emph{molecular dynamics (MD)}, which simulates molecular trajectories
by numerically integrating equations of motion.
The resulting trajectories record both the variation within physical states and the transitions between them.
Extracting the corresponding kinetic information requires \emph{molecular descriptors} that
resolve relevant structural and dynamical differences.
Such descriptors enable analyses such as Markov state models (MSMs) \citep{Zwanzig.1983,Singhal.2004}, which describe kinetics
via transitions between discrete states.

Notably, conventional descriptors such as backbone torsions, pairwise distances, or contact maps are tied to a single system as their coordinates refer to specific residues or atom pairs and therefore do not carry the same meaning across different proteins.
However, recent MD datasets such as \mdcath\ \citep{Mirarchi.2024}, \texttt{ATLAS} \citep{VanderMeersche.2024} or \texttt{GPCRmd} \citep{Rodriguez.2020} span simulations of thousands of proteins and thereby invite learning a \emph{shared} representation space.
Such shared embeddings are available from protein language models (PLMs) trained as foundation models on a large and diverse corpus of data.
Many popular PLMs \citep[e.g.,][]{Elnaggar.2022,Lin.2023} operate directly on the level of amino acid sequences and therefore cannot distinguish between different conformations of the same protein.
Structure-aware extensions  \citep[e.g.,][]{Hsu.2022,Su.2024} can discern them, but are typically trained on a single near-native state, rather than on the many conformations a protein samples in MD simulations.
Closer to our setting, \cite{Aboelsoud.2026} add MD trajectories as a modality to the protein foundation model OneProt \citep{Floegge.2025}, but pool each trajectory into a single embedding, summarizing a protein's variability rather than representing individual conformations.
Still, all these approaches rely on learning representations via large-scale pretraining, and their massive compute requirements motivate the question:
\emph{How well can principled, conformation-specific descriptors
represent protein structure and dynamics without large-scale pretraining?}

To obtain descriptors suitable for learning across protein domains, we build on \emph{persistent homology (PH)} \citep{Edelsbrunner00a, Zomorodian05a}, i.e., a tool from topological data analysis that summarizes shape across spatial scales. PH has proven informative for analyzing, classifying, and predicting properties of proteins \citep{Xia.2014,cang2015classification,cang2017topologynet}, and for distinguishing conformations along folding trajectories \citep{Ichinomiya.2020,Ichinomiya.2022}. Upon vectorization, these topological summaries provide a common representation space without requiring feature correspondence between proteins.

Motivated by these developments as well as prior work on incorporating biological information into PH-based analyses \citep{cang2018representability,dey2018classification}, and the availability of large-scale MD datasets, we seek to extend this computational pipeline to large collections of proteins and their trajectories by computing descriptors that are compatible across domains and informed by biological knowledge that emphasizes relevant aspects of molecular geometry. Effectively, we ask: \emph{How well can biologically informed PH descriptors represent protein structure and dynamics while supporting learning across
proteins and generalization to unseen domains?}

\vspace{-1ex}

\paragraph{Contribution} To investigate the aforementioned questions and to address the challenge of designing topologically grounded representations of protein structure and dynamics, we first introduce the \emph{masked Flood complex} as an adaptation of a recent method for efficient PH computation tailored to the characteristics of proteins. We then probe representation quality of the topological summaries by empirically showing that they carry important information about global shape and capture instantaneous and local variations. Further, we assess the quality of topologically informed MSMs to probe representations for their kinetic content. Finally, when used within the recent \marsfm\ model of \cite{Kapusniak.2026}, where these MSMs guide the training of a
generative model of molecular structures, we show (1) consistently strong performance across a range of ensemble statistics on the \mdcath\ dataset and (2) promising transfer to the regime of fast folding proteins \citep{Lindorff.2011} which fundamentally differs from the \mdcath\ training data.

\begin{figure}[t!]
\centering{
\includegraphics[width=1.0\textwidth]{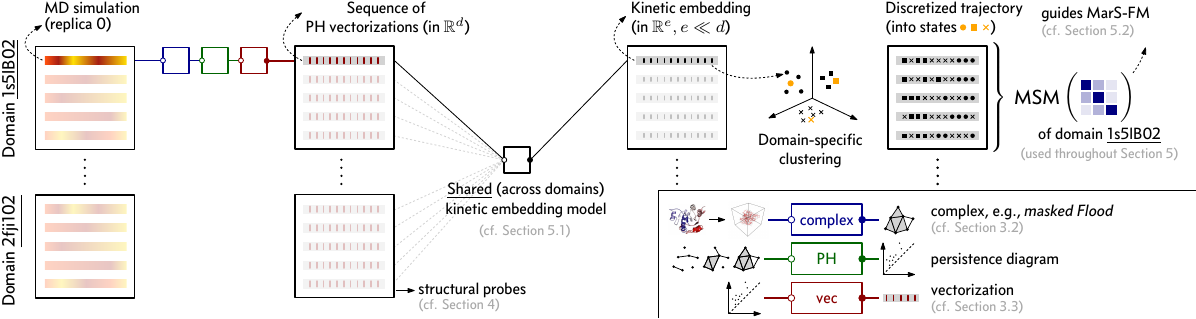}}
\caption{\label{fig:intro} For each frame of an MD trajectory (here, one replica of the \mdcath\ domain 1s5lB02), we compute PH, e.g., via our masked Flood complex (\cref{subsection:maskedflooder}), and vectorize the persistence diagrams (\cref{subsection:vectorization}), turning the trajectory into a sequence of PH vectorizations. A kinetic embedding model shared across domains maps these to a low-dimensional space, in which domain-specific clustering yields discretized trajectories and an MSM per domain (\cref{section:dynamics}). We probe the PH vectorizations for structural information (\cref{section:probes}) and evaluate the MSMs directly (\cref{subsection:msmlearning}) and as guidance for generative modeling (\cref{subsection:generativemodeling}). The legend (bottom right) details the per-frame steps: the complex construction yields a nested family of complexes over increasing radius $r$, i.e., a filtration, and PH summarizes how their topology changes along it in a persistence diagram, which is then vectorized.}
\vspace{-0.2cm}
\end{figure}

\vfill
\section{Related Work}
\label{section:relatedwork}

\vspace{-0.1cm}
\paragraph{Computational topology for molecular structure and dynamics}
Prior work has used PH-based descriptors to characterize protein
structure and flexibility \citep{Xia.2014}, classify proteins
\citep{cang2015classification}, and predict molecular properties
\citep{Cang.2017,cang2017topologynet}.
More recent approaches combine \emph{local} atomic PH with graph learning
for interface-quality assessment \citep{han2025topoqa}, or analyze
persistence generators across large structural databases
\citep{madsen2025universe}.
Biological information already enters these pipelines through
element selection, interaction-dependent dissimilarities
\citep{cang2018representability}, and constructions informed by
atomic radii and bonds \citep{dey2018classification}.
In \cref{def:mflood}, we will see that our approach builds on this principle through a different mechanism: residue-dependent masks restrict the atoms permitted to cover each simplex in the landmark-based simplicial complex construction of \cite{Graf.2026}.

In the context of MD simulations, PH has been used to study folding via representative cycles and residue-pair features, followed by dimensionality reduction \citep{Ichinomiya.2020,Ichinomiya.2022}.
While these features aid structural interpretation,
their coordinates depend on protein-specific residue indexing.
Our topological summary vectorization enables learning across proteins of different sizes without residue correspondence, while kinetic
coordinates are learned from time-lagged statistics.
A complementary line of work analyzes conformational state-space topology: persistent clustering uses kinetic distances to
identify metastable states \citep{chang2013metastable}, and
Mapper-based methods characterize pathways
\citep{Yao.2009,Donati.2025}. In contrast, we apply PH to individual conformations and use the resulting descriptors downstream, e.g., to learn a shared kinetic representation across domains, followed by domain-specific
MSM estimation.

\vspace{-0.2ex}
\paragraph{Learning from protein structure} Pretrained encoders map sequences \citep{Lin.2023} or structures \citep{Hsu.2022,Su.2024} to embeddings reused across tasks; ESM-IF1, for instance, is a transformer with GVP input layers \citep{Jing.2021}, trained for inverse folding on millions of predicted structures. Equivariant networks such as EGNN \citep{Satorras.2021} and GotenNet \citep{Aykent.2025} instead usually learn from coordinates end-to-end for each task. Our descriptors are fixed before any learning, with biological priors entering through the simplicial complex, and are read out by simple probes.

\vspace{-0.2ex}
\paragraph{Generative modeling of molecular ensembles}
While learning-based methods
have a long history in MD \citep{Blank.1995}, most prominently through learned interatomic potentials that amortize the cost of evaluating atomic interactions \citep{Unke.2021, Friederich.2021}, our work primarily connects to generative approaches for sampling molecular geometries \citep{Noe.2019,Lewis.2025,Plainer.2025,Costa.2025,
Bigi.2026,Daigavane.2026}.
Temporal information can enter through models of conformational
transitions \citep{Jing.2024} or through MSM-guided selection
of training pairs \citep{Kapusniak.2026}.
Our downstream evaluation builds on the latter approach where
we replace the MSMs used in \marsfm\ with those learned from
topological descriptors and assess the resulting generated
ensembles (\cref{subsection:generativemodeling}).
This tests whether the kinetic information captured by our
representations benefits generative-model training.

\vspace{-0.3ex}
\section{Method}
\label{section:method}
The blueprint of our approach is to map atomic configurations to vectorized topological summaries from PH. Furthermore, we introduce a protein-structure–biased PH variant by constructing a simplicial complex that highlights relevant residue interactions. Details are in \cref{subsection:maskedflooder,subsection:vectorization}.

\vspace{-0.2ex}
\subsection{Persistent homology primer}
\label{subsection:ph}
One way to extract topological features from a point cloud $X\subset \mathbb R^3$ is persistent homology. When computing PH, one represents the point cloud as a family of combinatorial objects $\{\Sigma_r\colon r\ge 0\}$ called \emph{simplicial complexes}, each corresponding to a radius or scale $r>0$ at which the point cloud is studied. PH then tracks the changes in the structure of $\Sigma_r$, and therefore in the underlying point cloud $X$ as the scale varies, and summarizes them in persistence diagrams. More precisely, a simplicial complex $\Sigma$ over the set $X$ is a set of subsets $\sigma\subset X$, with the property that if $\sigma \in \Sigma$ then also $\tau \in \Sigma$ for all $\tau \subset \sigma$.  Moreover, computing PH requires that the family $\{\Sigma_r\}_r$ of simplicial complexes satisfies $\Sigma_s \subset \Sigma_r$ whenever $s\le r$, i.e., it needs to be a 
\emph{filtration}.

Typically, for each simplex $\sigma \in \Sigma$ there is a minimal $r$ such that $\sigma \in \Sigma_r$, i.e., the moment when it appears.
This value $r$ is called the filtration value of $\sigma$ and a function $f:\Sigma\to \mathbb R$ that assigns to each simplex a filtration value is called a filtration function. 
Conversely, any function $f:\Sigma \to \mathbb R$ that satisfies $f(\tau)\le f(\sigma)$ for every $\tau,\sigma\in \Sigma$ with $\tau\subset \sigma$ induces a well-defined filtration on $\Sigma$ in terms of the sublevel sets $\Sigma_r = \{\sigma \in \Sigma \colon f(\sigma) \le r\}$. Topological information about the filtration can be summarized in multisets of tuples $(b,d)\in \mathbb{R}^2$ called persistence diagrams. Each tuple corresponds to a topological feature; its entries specify at which filtration values $r$ the feature appears (is born) and vanishes (dies), resp.
For $X\subset \mathbb R^3$, there are typically three persistence diagrams of interest; $H_0$ corresponding to connected components, $H_1$ to loops, and $H_2$ to voids.

For commonly used filtrations, the appearance of a simplex in a complex $\Sigma_r$ means that its vertices are close relative to $r$. For example, the Vietoris-Rips complex consists of all subsets of $X$ with diameter at most $r$,
and the Alpha complex consists of simplices in the Delaunay triangulation of $X$ whose vertices lie on the boundary of a ball of radius at most $r$ containing no other points of $X$.

For protein data with atoms represented as points in Euclidean space, however, not all points are equivalent: the residues' $\alpha$-carbons (C$_\alpha$) form a natural set of distinguished atoms. Moreover, atoms within the same residue remain in close spatial proximity, and their pairwise distances are typically less affected by non-trivial transitions such as protein folding or the opening of cryptic pockets.
Incorporating this structural prior motivates the \emph{masked Flood complex}. %

\vskip-2ex
\subsection{Masked Flood complex}
\label{subsection:maskedflooder}

Recently, \cite{Graf.2026} introduced Flood complexes, a class of simplicial complexes that allow for efficient computation of PH of large-scale point clouds. We restate their definition below.

\begin{definition}
    Let $X,L\subset \mathbb R^d$ be finite sets. The Flood complex $\Flood_r(X,L)$ at radius $r\ge 0$ is %
    \begin{equation}
    \label{eqn:floodcomplex}
        \Flood_r(X,L) =  \left \{ \sigma \in \Del(L) \colon \hull(\sigma) \subset \bigcup_{x\in X} B_r(x) \right \} \enspace.
    \end{equation}
    \vspace{-0.3cm}
\end{definition}
The set $X$ is the point cloud of interest whereas $L$ is typically a small subset of $X$. As $\Flood_r(X,L)$ is a subcomplex of the Delaunay triangulation $\Del(L)$, its number of simplices is comparably small, allowing for fast PH computation, while the filtration values are still informed by the full point cloud $X$ due to the requirement that a simplex $\sigma$ is in $\Flood_r(X,L)$ only if its convex hull $\hull (\sigma)$ is covered, or ``flooded'' by the union of closed balls $\bigcup_{x\in X} B_r(x)$.
Its filtration values can be computed efficiently in parallel on a GPU, with further speedups from precomputing, for each simplex $\sigma$, a candidate mask of points $x\in X$ on which the simplex's filtration value must be realized.

While originally introduced as a means for efficient PH computation, both landmark selection and masking can be used to \emph{emphasize specific properties of the point cloud} in the resulting filtration and its persistent homology.
The definition of the masked Flood complex below captures this very idea.

\vskip1ex
\begin{definition}
\label{def:mflood}
    Let $L\subset \mathbb R^d$ be a finite set, and for each $\sigma \in \operatorname{Del}(L)$ let $X_\sigma \subset \mathbb R^d$ be finite such that $X_\sigma \subset X_\tau$ for all $\tau \subset \sigma$.
    The masked Flood complex at radius $r\ge0$ is the simplicial complex
    \begin{equation}
        \mFlood_r(\{X_\sigma\}_{\sigma\in \operatorname{Del}(L)},L) = \Big\{\sigma \in \operatorname{Del}(L) \colon  \hull{(\sigma)} \subset \bigcup_{x\in X_\sigma} B_r(x) \Big\}\enspace.
            \vspace{-0.2cm}
    \end{equation}
\end{definition}
In practice, we have some point cloud $X$ and a distinguished set of landmarks, and we assign to each simplex $\sigma\in \Del(L)$ a set $X_\sigma \subset X$, i.e., a mask that specifies which points of $X$ contribute to the filtration value of $\sigma$.
By construction, the masked Flood complex is then a subcomplex of the original Flood complex from \cite{Graf.2026} on $X$, and they agree if $X_\sigma = X$ for all $\sigma \in \operatorname{Del}(L)$.
We include the requirement that $\tau \subset \sigma$ implies $X_\tau \supset X_\sigma$ in \cref{def:mflood} as it ensures that at each radius $r$ the masked Flood complex is a well-defined simplicial complex and therefore induces a filtration of $\operatorname{Del}(L)$.
Indeed, denoting by $d$ the Euclidean distance, the filtration value $f(\sigma)$ of each simplex $\sigma \in \Del(L)$ is $f(\sigma) = \max_{p\in \hull(\sigma)} \min_{x\in X_{\sigma}} d(p,x)$. Hence $\tau \subset \sigma$ implies $f(\tau) \le f(\sigma)$, since
\begin{align}
    \label{eq:filtration_good}
        f(\tau)
         = \max_{p\in \hull(\tau)} \min_{x\in X_{\tau}} d(p,x)
         \le \max_{p\in \hull(\sigma)} \min_{x\in X_{\tau}} d(p,x)
         \le \max_{p \in \hull(\sigma)} \min_{x\in X_{\sigma}} d(p,x)
         = f(\sigma)
        \enspace.
\end{align}
Following \cite{Graf.2026}, we approximate filtration values using a grid on the simplices' convex hulls.
In \cref{section:app:theory} we verify that this still results in a well-defined filtration in the presence of masking, and, for fixed landmark positions, bottleneck stability with respect to the masks.

\begin{figure}
    \centering{
    \includegraphics[width=.95\textwidth]{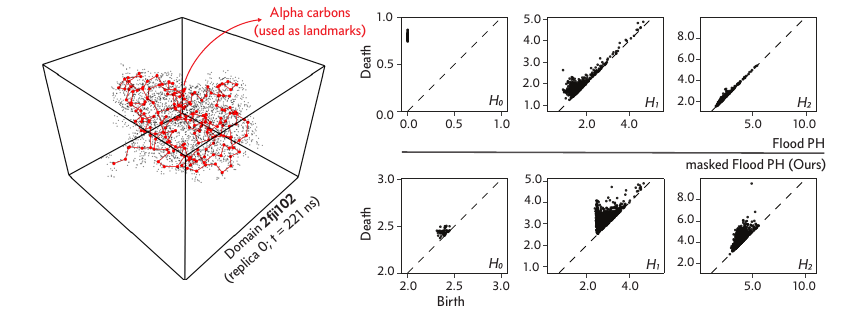}}
    \vspace{-0.2cm}
    \caption{\textbf{Left}: Point cloud corresponding to a folded protein. Highlighted in red are the C$_\alpha$ atoms.\\ \textbf{Right}: Persistence diagrams produced by Flood PH (top) and masked Flood PH (bottom).}
    \label{fig:ph_compare}
    \vspace{-0.4cm}
\end{figure}
For proteins, we use their C$_\alpha$ atoms as landmarks, i.e., simplices are sets of C$_\alpha$ atoms, and therefore correspond to sets of residues. To emphasize spatial relationships across residues, we design the masks as follows: For each vertex $l\in L$, its mask $X_l$ consists only of atoms that are not part of the same residue. For higher-dim. simplices, we take the intersection of their vertex masks, i.e., the largest possible set of atoms that is compatible with the nesting condition for the masks from \cref{def:mflood}. Thus, a simplex is flooded only by atoms from residues not represented by its vertices, e.g., an edge is flooded only by atoms from outside the two residues whose C$_\alpha$ atoms it connects.

\begin{wraptable}{r}{0.42\textwidth}
\vspace{-0.45cm}
\caption{Wall time of PH computation ($H_0$--$H_2$) per persistence diagram (mean $\pm$ std over 500 \mdcath\ domains) and estimated total for all \mdcath\ conformations; cf.  \cref{tab:performance}.\label{tab:runtime}}
\vspace{-0.25cm}
\centering
\begin{small}
\begin{tabular}{lrr}
\toprule
 & \makecell{Per diagram\\(ms)} & \makecell{Total\\(h)} \\
\midrule
\floodph &  14.4 \std{\phantom{0}7.1} & 46 \\
\mfloodph & 28.4 \std{21.1} & 91 \\
\alphaph &  94.1 \std{63.5} & 301 \\
\bottomrule
\end{tabular}
\end{small}
\vspace{-0.3cm}
\end{wraptable}
\cref{fig:ph_compare} compares the resulting persistence diagrams to those from the Flood complex. The differences are most transparent in $H_0$ where, in contrast to most simplicial complexes, the features are \emph{not} all born at radius $r=0$, but at differing positive radii, because landmarks $l$ are born at $\min_{x\in X_l} d(l, x)$, as $l \not \in X_l$.
Moreover, there are fewer points in $H_0$ than there are landmarks, as a vertex can appear at the same time as an edge connecting it to an existing connected component, and thus, without creating a new one (see \cref{fig:masked_flood_edge} in the appendix for an illustration). We also observe that more points in the $H_1$ and $H_2$ diagrams are persistent, i.e., well-separated from the diagonal.
Using masks to encode structural information rather than solely for efficiency doubles the runtime relative to Flood PH. Still, masked Flood PH is more than three times faster than all-atom \alphaph, since the complex is built only on the C$_\alpha$ landmarks and filtration values are computed in parallel on the GPU (see \cref{tab:runtime,sec:runtime}).

\vspace{-0.2cm}
\subsection{Vectorizing persistence diagrams}
\label{subsection:vectorization}
To use persistence diagrams in learning tasks, we first need to represent them as vectors. In particular, to enable learning across molecular configurations, we require a map into a coordinate system consistent across frames and MD trajectories. To this end, we adopt a vectorization based on exponential structure elements \citep{Hofer.2019}, which maps persistence diagrams (up to dimension 2) into a vector by using the same structure elements (fitted on training data) for all diagrams. 

\vskip1ex
\emph{In the following, we (1) probe the (PH + vectorization) representations for their structural information content (\cref{section:probes}), (2) assess whether they support learning low-dimensional kinetic embeddings shared across protein domains for MSM estimation (\cref{subsection:msmlearning}), and (3) use the MSMs to guide sampling of training pairs for a generative model (\cref{subsection:generativemodeling}).}

\vspace{-0.2cm}
\section{Probing structural representation quality}
\label{section:probes}
We  evaluate the extent to which vectorized persistence diagrams 
capture \emph{structurally relevant}
 information in molecular systems. Evaluation in terms of \emph{dynamic information} is done in \cref{section:dynamics}.
For comparison to masked Flood PH (\mfloodph), we evaluate PH (up to dim. 2) computed from Alpha complexes on the full point cloud $X$ (\alphaph), Alpha \& Rips complexes on the C$_\alpha$ atoms (\alphaphca, \ripsphca), and Flood PH \citep{Graf.2026} with C$_{\alpha}$ 
atoms as landmarks (\floodph).

\vskip-0.2ex
\paragraph{Datasets}
We rely on the \mdcath\ \citep{Mirarchi.2024} dataset, inheriting a train-val-test split from \cite{Kapusniak.2026} that enforces strict sequence dissimilarity between train and test domains. \mdcath\  consists of MD trajectories from 5,398 domains. For each domain, we use the five simulation replicas at 450 K; most trajectories are 500 ns long, with coordinates recorded every 1 ns.

\vspace{-0.2ex}
\paragraph{Implementation}
To isolate representational quality, we probe the vectorized persistence diagrams with simple models: a single linear layer, and 2-layer MLPs with ELU activations. Persistence diagrams are vectorized using 128 exponential structure elements \citep{Hofer.2019} per $H_0$, $H_1$ and $H_2$ diagram, yielding a 384-dim. representation.
We refer to \cref{app:sec:vectorization} for a description of the vectorization method and to \cref{app:sec:tasks_setup} for architectural and training details.

We probe two complementary aspects of representation quality:
\emph{First}, we evaluate protein class prediction at the replica level, assessing whether time-aggregated representations retain global structural discriminability.
\emph{Second}, we regress physical observables,
testing whether the representations capture instantaneous conformational variability.
Specifically, we seek to predict the radius of gyration (\texttt{RG}), the secondary structure fraction (\texttt{SSF}),
the fraction of native contacts (\texttt{FNC}), and the root mean-square deviation (\texttt{RMSD})
to the replica's starting frame (see \cref{section:app:observablesdef} for their definition).

\vspace{-0.5ex}
\paragraph{Baselines}
As a \emph{physical} \emph{baseline}, we encode protein domains as (num.~residues, 7)-shaped sequences, with residues represented by the sines and cosines of their $\phi$- and $\psi$-torsion angles, and the 3D coordinates of their $\alpha$-carbons. On these sequences, we train a 1D convolutional network with windowing along the residue axis. This respects the sequential nature of proteins and allows handling different domain sizes. For architectural details of the baseline and our approach see \cref{app:sec:tasks_setup}.

As \emph{foundation model baselines}, we include \texttt{ESM2} \citep{Lin.2023} and the inverse folding model \texttt{ESM-IF1} \citep{Hsu.2022}.
Their pretraining data, UniRef sequences and AlphaFold2 predictions of such sequences, may include our test domains, which cannot be verified.
It can for the CATH training split of \texttt{ESM-IF1}: 53.5\% of our test domains are identical or near-identical to one of its structures (\cref{sec:app:esmoverlap}), so we gray out all results involving \texttt{ESM-IF1}.

As \emph{learned end-to-end baselines} providing strong task-specific performance references, we use \texttt{EGNN} \citep{Satorras.2021} and \texttt{GotenNet} \citep{Aykent.2025} trained directly on the downstream tasks.

\vspace{-0.5ex}
\paragraph{Classification of top-level protein class} 

\begin{wraptable}{r}{0.45\textwidth}
\vspace{-0.45cm}
\caption{Per-replica classification of CATH class. Balanced accuracy (mean $\pm$ std over 3 runs) in \%. \textbf{Best}/\underline{second-best} result excluding \texttt{ESM-IF1}.\label{table:clf}}
\vspace{-0.3em}
\centering{
\begin{small}
\begin{tabular}{lcc}
\toprule
            & Linear & MLP \\ %
\midrule
\mfloodph  & \underline{79.1} \std{0.7}	  & \textbf{81.2} \std{2.1} \\
\floodph             & 78.4 \std{2.0}                & \underline{80.0} \std{1.6} \\
\alphaphca           & 78.4 \std{1.0}	              & 78.2 \std{5.8} \\
\alphaph             & 71.3 \std{3.4}	              & 74.5 \std{3.1} \\
\ripsphca	         & \textbf{79.5} \std{0.8}        & 77.4 \std{5.3} \\
\midrule
(\rg, \ssf, \fnc, \rmsd)    & 54.3 \std{0.6} &  60.5 \std{0.4} \\
($\phi$, $\psi$) + C$_\alpha$ coords.  & -- & 70.8 \std{0.1}  \\
\midrule
 \texttt{ESM2\_35m} & \textbf{79.5} \std{0.2} & 77.3 \std{0.6} \\
\textcolor{lightgray}{\texttt{ESM-IF1}} & \textcolor{lightgray}{82.7} \std{1.8} & \textcolor{lightgray}{78.7} \std{3.6} \\
\textcolor{lightgray}{\texttt{(ESM2\_35m, ESM-IF1)}} & \textcolor{lightgray}{81.5} \std{1.6} & \textcolor{lightgray}{78.8} \std{1.8} \\
\midrule
\texttt{EGNN} & -- & 71.8 \std{5.1} \\
\texttt{GotenNet} & -- & 75.5 \std{7.9} \\
\bottomrule
\end{tabular}
\end{small}}
\vspace{-0.4cm}
\end{wraptable}

\begin{figure}[t!]
\centering{
\includegraphics[width=0.95\textwidth]{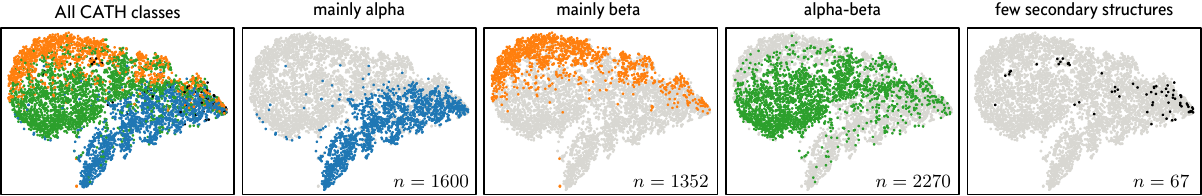}}
\caption{\label{fig:tsne} t-SNE embeddings of mean-pooled \mfloodph\ vectorizations of the 450 K \mdcath\ trajectories.}
\vspace{-0.3cm}
\end{figure}

We consider protein classification at the trajectory level, by assigning each MD simulation replica of a protein domain to one of four CATH classes (mainly alpha, mainly beta, alpha-beta and few secondary structures).
We obtain fixed-length features by averaging vectorized persistence diagrams over time.
The 2D t-SNE embeddings in \cref{fig:tsne} already suggest good class separability for the \mfloodph\ representations.

For a quantitative analysis, we train a linear classifier and a small MLP and report balanced accuracy in \cref{table:clf}. In addition to the aforementioned baselines, we also include a comparison with a feature set based on four physical observables (\rg, \ssf, \fnc, \rmsd).

All PH-based representations retain more class information than the physical references and, for most, at least as much as the end-to-end models.
Compared with foundation-model representations, we see that linear probes of \ripsphca\ and \mfloodph\ match those of \texttt{ESM2\_35m}, the best of four \texttt{ESM2} sizes (\cref{table:app:clf}), and that \mfloodph\ exceeds them when probed by the same MLP.
The highest accuracy is obtained from \texttt{ESM-IF1} representations with a linear probe, within one standard deviation of MLP probes of \mfloodph.
\emph{Overall}, topological summaries capture global conformational information beyond simple physical descriptors, on par with end-to-end trained models and learned representations (despite their advantage on the test split).

\vspace{-0.2cm}
\paragraph{Regression of physical observables}
We probe how well common structural descriptors (\rg, \ssf, \rmsd, \fnc) can be read out from a representation.
These observables are standard in the analysis of MD simulations and describe structure at complementary scales: \ssf\ reflects local secondary structure, \fnc\ the preservation of native tertiary contacts \citep{Best.2013}, and \rg\ and \rmsd\ global compactness and overall deviation from the starting conformation.

For each target, we train a separate scalar regression model on all time points and all replicas of the training domains and evaluate across all time points of all replicas of the test domains. \cref{table:regression} lists the coefficient of determination ($R^2$) across all targets.

\begin{table}[h]
\vspace{-0.1cm}
\caption{Regression of physical observables per time point across all testing domains of \mdcath. Reported is the mean coefficient of determination, $R^2$ ↑, computed over training runs of the linear and MLP model from different random seeds. Standard deviations are at most 0.005 for fixed representations and up to 0.026 for end-to-end models (\cref{table:app:regression}).
\textbf{Best}/\underline{second-best} PH-based result. \label{table:regression}}
\vspace{-0.2cm}
\centering{
\begin{small}
\begin{tabular}{l *{8}{S[table-format=1.3]}}
\toprule
 & \multicolumn{4}{c}{Linear} & \multicolumn{4}{c}{MLP} \\
\cmidrule(lr){2-5} \cmidrule(lr){6-9}
 & {\fnc} & {\rg} & {\rmsd} & {\ssf}
 & {\fnc} & {\rg} & {\rmsd} & {\ssf} \\
\midrule
\mfloodph
    & \best{0.691} & \second{0.728} & \second{0.683} & \second{0.652} & \second{0.860} & 0.861 & 0.781 & 0.842 \\
\floodph
    & \second{0.686} & 0.719 & 0.675 & \best{0.658} & \best{0.863} & \second{0.864} & \second{0.782} & \second{0.853} \\
\alphaph
    & 0.668 & \best{0.831} & \best{0.708} & 0.623 & 0.858 & \best{0.921} & \best{0.798} & 0.815 \\
\alphaphca
    & 0.669 & 0.675 & 0.656 & 0.650 & 0.859 & 0.849 & 0.772 & \best{0.854} \\
\ripsphca
    & 0.630 & 0.559 & 0.597 & 0.599 & 0.840 & 0.772 & 0.732 & 0.828 \\
\midrule
{($\phi$, $\psi$) + C$_\alpha$ coords.}  &  {--} & {--} & {--} & {--} & 0.762 & 0.332 & 0.578 & 0.841 \\
\midrule
\textcolor{lightgray}{\texttt{ESM-IF1}}  &  \textcolor{lightgray}{0.918} & \textcolor{lightgray}{0.904} & \textcolor{lightgray}{0.822} & \textcolor{lightgray}{0.933} & \textcolor{lightgray}{0.929} & \textcolor{lightgray}{0.946} & \textcolor{lightgray}{0.849} & \textcolor{lightgray}{0.934} \\
\textcolor{lightgray}{\texttt{(ESM2\_35m, ESM-IF1)}}  &  \textcolor{lightgray}{0.929} & \textcolor{lightgray}{0.905} & \textcolor{lightgray}{0.829} & \textcolor{lightgray}{0.932} & \textcolor{lightgray}{0.942} & \textcolor{lightgray}{0.944} & \textcolor{lightgray}{0.853} & \textcolor{lightgray}{0.934} \\
\midrule
\texttt{EGNN}  &  {--} & {--} & {--} & {--} & 0.878 & 0.955 & 0.799 & 0.863 \\
\texttt{GotenNet}  &  {--} & {--} & {--} & {--} & 0.879 & 0.953 & 0.808 & 0.868 \\
\bottomrule
\end{tabular}
\end{small}}
\vspace{-0.1cm}
\end{table}

Most targets are recovered better from PH-based representations than from the physical reference. End-to-end GNNs recover \rg\ best, and pretrained \texttt{ESM-IF1} 
and its fusion with \texttt{ESM2\_35m} excel in recovering the remaining observables, in line with their task-specific training and scale.

\vspace{-0.5ex}
\section{Dynamics}
\label{section:dynamics}
\vspace{-0.5ex}
We assess the utility of vectorized persistence diagrams in a series of experiments building upon each other. We first use them to fit MSMs on \mdcath, which we then use as a drop-in replacement for the ones originally used (based on physical observables) to guide the training of \marsfm. Finally, we study the transferability of the obtained \marsfm\ model to fast folding proteins (via finetuning).

\vspace{-0.5ex}
\subsection{Learning a shared kinetic embedding space and MSM estimation}
\label{subsection:msmlearning}
To study the dynamic information captured by vectorized persistence diagrams methodically, 
we rely on MSMs to assess the capability of features to resolve kinetically relevant processes;
typically the slowest modes of a dynamical system, often associated with rare but meaningful transitions \citep{Prinz.2011}.
To construct an MSM, conformations are usually described by physical observables, or projected onto the directions of
slowest decorrelation via TICA \citep{Molgedey.1994, Schwantes.2013, Perez.2013}.
The resulting representations are then discretized by clustering, optionally followed by coarse-graining \citep{Roblitz.2013}.
Usually, this workflow is performed on one or more MD trajectories of a \emph{single system},
and the resulting MSM quality is tied to the amount and quality of the available data, as well as the chosen descriptors. The availability of datasets comprising trajectories from diverse protein domains motivates leveraging population-level information by learning a shared representation space \emph{across domains}. 

To this end, we adapt the VAMPNet framework \citep{Mardt.2018} to train a shared model across systems (cf.~\cref{subsection:app:vampnet}).
After training, we map all trajectory frames to the shared latent space and, for each domain, discretize the representations of each replica into $k$ states using $k$-means++ \citep{Arthur.2007}. The resulting discrete trajectories are then used for MSM estimation. 

\vspace{-1.2ex}
\paragraph{Implementation}
As input for the kinetic embedding we use the PH vectorizations of \cref{section:probes} based on the 450 K simulations of \mdcath. 
As VAMPNet, we use a simple 5-layer MLP with a 7D output (for training details, see \cref{subsection:app:vampnet}).
MSMs are estimated per domain by maximum likelihood with a reversible transition matrix from effective transition counts at lag 50 ns, restricted to the largest connected set.\footnote{Although individual \mdcath\ trajectories are short ($\leq 500$~ns), they suffice for valid MSMs (\cref{subsubsection:app:mdcath:validmsm}).}
The choice of the lag time and the temperature for MD simulations is determined by the subsequent generative modeling experiment for which we reuse the MSMs, as we want to preserve the setup from \cite{Kapusniak.2026} for a sound comparison.

\vspace{-1.2ex}
\paragraph{Quality measures}
To assess the quality of the produced MSMs, we use three complementary quality measures: \texttt{vamp-2}, which measures how well the slow processes in the data are resolved, \texttt{CK RMSE} computed from Chapman–Kolmogorov (CK) tests which measure whether the resulting state decomposition yields Markovian dynamics at the chosen lag time, and $\pi$-\texttt{JS}, the Jensen–Shannon divergence between the MSM-implied stationary distribution and the empirical distribution. As we reserve the testing portion of \mdcath\ for evaluating the generative model in \cref{subsection:generativemodeling}, we report scores on its validation split.
For each domain, we fit MSMs on four of the five MD replicas and compute the \texttt{vamp-2} and $\pi$-\texttt{JS} on the held-out replica. The resulting five scores are averaged per domain, and \cref{tab:msm-quality} reports the mean and standard deviation across domains.
Due to the short \mdcath\ trajectories, cross-validating the CK test is unstable
(see \hyperref[par:app:ck]{Appendix B}). In any case, it stands as a valid self-consistency check for Markovianity, rather than a generalization measure.

\vspace{-1.2ex}
\paragraph{Baselines}
We compare against MSMs fitted on (i) representations from a convolutional VAMPNet trained on backbone torsion angles ($\phi$, $\psi$) and C$_\alpha$ coordinates, (ii) the physical observables
(\rg, \ssf), as in \citet{Kapusniak.2026}, (iii) 7-dimensional TICA projections of commonly used descriptors: torsion angles (sine-cosine pairs of backbone and sidechain angles) and the solvent-accessible surface area (SASA) in the main part and other typically used descriptors in \cref{subsubsection:app:mdcath:msm}.

\vspace{-1.2ex}
\paragraph{Results}
\begin{wraptable}{r}{0.54\textwidth}
\vspace{-0.45cm}
 \caption{MSM quality measures at lag time ${\tau=50}$~ns with 7-dimensional VAMPNet features and 10 micro states. \texttt{CK RMSE} is evaluated at lag times 50 and 100~ns on 3 macro states. \textbf{Best}/\underline{second-best} overall.} \label{tab:msm-quality}
    \centering
    \vspace{-0.2cm}
    \begin{small}
    \setlength{\tabcolsep}{3pt}
\begin{tabular}{lccccccc}
    \toprule
    & \texttt{vamp-2} ↑ & \texttt{CK RMSE} ↓ & $\pi$-\texttt{JS} ↓ \\
    \midrule
    \mfloodph & 1.503 \std{.010} & \textbf{0.034} \std{.001}  & \underline{0.120} \std{.004}  \\
    \floodph            & 1.482 \std{.013}           & 0.051 \std{.002}                 & 0.219 \std{.006}  \\
    \alphaphca          & \underline{1.505} \std{.016}  & 0.053 \std{.002}                 & 0.246 \std{.005} \\
    \alphaph            & 1.463 \std{.010}           & \textbf{0.034} \std{.001}    & \textbf{0.109} \std{.004}  \\
    \ripsphca           & 1.477 \std{.014}           & 0.049 \std{.001} & 0.254 \std{.006} \\
    \midrule
    {($\phi$, $\psi$) + C$_\alpha$ coords.}
        & 1.113 \std{.004}
        & 0.046 \std{.001}
        & 0.332 \std{.004}  \\
    (\rg, \ssf) & \textbf{1.513} \std{.004} & 0.039 \std{.001} &  0.124 \std{.004}  \\
    \midrule
    \texttt{TICA (Torsion)} &  1.264 \std{.002} & 0.076 \std{.000} & 0.139 \std{.001} \\
    \texttt{TICA (SASA)} & 1.439 \std{.002} & \textbf{0.034} \std{.000} & 0.258 \std{.001} \\
    \bottomrule
    \end{tabular}
    \end{small}
    \vspace{-1.0cm}
\end{wraptable}
\mfloodph\ is the only representation on par with or better than (\rg, \ssf) on all three measures, although the latter are tailored to unfolding, the dominant slow process at 450~K.
Even with one embedding shared across domains, \mfloodph\ captures the slow processes as well as these hand-picked observables and yields more Markovian MSMs.
Masking is decisive: relative to \floodph, it lowers \texttt{CK RMSE} by a third and nearly halves $\pi$-\texttt{JS}.
\mfloodph-based MSMs could thus replace (\rg, \ssf) in \marsfm\ without loss in MSM quality.

\vskip-0.1ex

\vspace{0.15cm}
\subsection{Generative modeling}
\label{subsection:generativemodeling}

In \marsfm\ \citep{Kapusniak.2026}, a generative model learns by flow matching to produce a target frame from a source frame, with pairs drawn from MSM state transitions of each domain.
We replace its (\rg, \ssf)-based MSMs by PH-based ones to test whether the kinetic information of PH carries over to generation, on held-out \mdcath\ domains and, after finetuning, on fast folding proteins.

\vspace{-0.15cm}
\paragraph{mdCATH}
With all other settings of \cite{Kapusniak.2026} unchanged, we compare generated and MD ensembles in conformational diversity (pairwise RMSD), residue flexibility (RMSF), compactness (\rg), secondary structure (\ssf) and folding free energy ($\Delta G$), as defined in \cref{section:app:observablesdef}. All PH variants improve on the original \marsfm\ in every statistic of \cref{table:mainclean} except RG KL, where they are closer to the \texttt{MD oracle}. As the differences are often small, we also test their significance: five of the six significant differences favor \marsfmmflood\ (\cref{tab:app:gensignificance}; full results in \cref{subsection:app:mdcathgen}).

\begin{table}[h]
\caption{Ensemble statistics on the \mdcath\ test domains, mean \std{1 std. dev.} over 5 evaluation runs; \marsfm\ from its released checkpoint. \texttt{MD oracle}: one MD replica scored against all five; \marsfm\texttt{-random}: MSMs fitted to 2D Gaussian random features. \textbf{Best} result, excluding oracle and random.}
\label{table:mainclean}
\centering
\begin{small}
\begin{tabular}{l *{6}{c}}
\toprule
 & \makecell{Pairwise\\RMSD r $\uparrow$} & \makecell{Global\\RMSF r $\uparrow$} & RG KL $\downarrow$  & SSF KL $\downarrow$ & SSF JSD $\downarrow$ & $\Delta$G MAE $\downarrow$ \\
\midrule
\texttt{MD oracle}    & 0.82 \std{.014} & 0.82 \std{.012} & 0.58 \std{.020} & 0.61 \std{.048} & 0.07 \std{.003} & 0.90 \std{.046}\\
\texttt{\marsfm -\texttt{random}}     & 0.57 \std{.004} & 0.68 \std{.003} & 0.51 \std{.002} & 1.52 \std{.009}  & 0.18 \std{.001} & 1.36 \std{.007}\\
\midrule
\marsfm   & 0.64 \std{.002} & 0.71 \std{.002} & \textbf{0.55} \std{.002} & 1.05 \std{.007} & 0.14 \std{.000} & 0.99 \std{.009} \\
\marsfmmflood      & \textbf{0.68} \std{.003} & \textbf{0.74} \std{.001} & 0.58 \std{.003}  & 0.81 \std{.005} & \textbf{0.12} \std{.000} & 0.96 \std{.012} \\
\marsfmflood & 0.66 \std{.002} & 0.73 \std{.001} & 0.58 \std{.002} & 0.81 \std{.007} & \textbf{0.12} \std{.000}  & 0.97 \std{.011}  \\
\marsfmalphaca & 0.67 \std{.002} & 0.72 \std{.001} & 0.60 \std{.003}  & \textbf{0.78} \std{.006} & \textbf{0.12} \std{.000} & \textbf{0.91} \std{.004}  \\
\marsfmalpha & \textbf{0.68} \std{.004} & \textbf{0.74} \std{.002} & 0.59 \std{.002} & 0.85 \std{.007} & \textbf{0.12} \std{.000}  & 0.92 \std{.005}  \\

\bottomrule
\end{tabular}
\end{small}
\end{table}

\paragraph{Transfer to fast folding proteins}
\begin{wraptable}{r}{0.55\textwidth}
\caption{Structural validation of ``fast folders''. Metrics are averaged over all
reference (MD), resp. generated, frames.
 \label{table:ff_validation}}
\vspace{-0.10cm}
\centering
\begin{small}
\setlength{\tabcolsep}{5pt}
\begin{tabular}{l *{3}{c}}
\toprule
 & \makecell{Bond \\ RMSZ $\downarrow$} & \makecell{Angle \\RMSZ $\downarrow$} &  \makecell{$\Delta G_\mathrm{fold}$ \\ \scriptsize{(kcal/mol)}} \\
\midrule
\texttt{MD reference} &  1.96 \std{0.02} & 1.86 \std{0.04} &  1.45 \std{0.67} \\
\midrule

\marsfm & 5.78 \std{0.89} & 3.21 \std{0.17} & 4.90 \std{1.01} \\
\marsfmmflood & 4.96 \std{0.74} & 3.10 \std{0.22} & 4.62 \std{0.81} \\
\midrule
\marsfm$^\text{\texttt{ft}}$ & 3.72 \std{0.53} & 2.44 \std{0.15} & 2.95 \std{1.03} \\
\marsfmmflood$^\text{\texttt{ft}}$ & 3.16 \std{0.28} & 2.36 \std{0.16} & 2.79 \std{0.96} \\
\bottomrule
\end{tabular}
\end{small}
\vspace{-0.5em}
\end{wraptable}
While \mdcath\ covers a broad structural space, its trajectories are too short to resolve asymptotic behavior. To address this limitation, following \citet{Costa.2025}, we assess \marsfmmflood\ on the long-timescale trajectories of \cite{Majewski.2023}, which provide $\sim$7 ms of aggregate simulation time for 12 fast folding proteins \citep{Lindorff.2011}.
Compared to \mdcath, these proteins are much smaller (10--80 residues), simulated at 350 K, and exhibit strong folding propensity (``fast folders''), representing the opposite of the unfolding dynamics within the 450 K \mdcath\ simulations.
To account for this disparity \emph{without} training on the fast folders, we finetune (\texttt{ft}) our \marsfmmflood\ model on the 348 K subset of \mdcath,
reduce lag time to 10 ns, and modify the sampling procedure by drawing the starting-frame cluster as one of the target rather than the
source conformation.
We obtain native-state conformations for \emph{all} 12 fast folding proteins (see \cref{tab:rmsd_fnc_gdt}) and promising structural validation measures (\cref{table:ff_validation}): RMSZ deviations in covalent backbone geometry, comparing to Engh-Huber bond and angle parameters \citep{Engh.1991}, and folding free energy $\Delta G_{\mathrm{fold}}$.
Covalent geometry is of similar quality for all proteins, including the smallest ones, for which \citet{Costa.2025} report chemically invalid states (\cref{subsection:app:fastfolders}).

\section{Discussion}
\label{section:discussion}

We set out to assess whether principled topological descriptors can serve as general-purpose representations of protein structure and dynamics without large-scale pretraining. To this end, we introduced the masked Flood complex, a protein-tailored simplicial complex that emphasizes inter-residue structure. Its vectorized persistence diagrams then place conformations of different proteins in a shared space. Our probing results show that they recover both global fold and per-frame conformational variability at a level comparable to learned protein representations. MSMs built on our descriptors improve ensemble statistics when used within \marsfm, and the resulting generative model transfers to fast folding proteins after finetuning. While our results also show that no PH variant is best across all probes and downstream tasks (e.g., \texttt{Alpha PH} on all atoms performing well in many cases), the masked Flood complex (\texttt{mFlood}) is the only construction among the strongest in every evaluation, which makes it the most consistent PH representation overall (\cref{tab:app:ranks}). Furthermore, computation time stays within a third of the time of all-atom \alphaph\ (\cref{tab:runtime}).
Future work could design masks for higher-order simplices systematically and summarize trajectory segments topologically, so that kinetic information enters the representation directly rather than only through a learned embedding.
Overall, masked Flood PH provides a computationally efficient means to incorporate prior knowledge into the construction of a simplicial complex whenever structure can be encoded as masks.

\clearpage
\section*{Acknowledgements}
This research was funded in whole or in part by the Austrian Science Fund (FWF) Grant-DOI 10.55776/DFH4791124.

\bibliographystyle{iclr2027_conference}
\bibliography{references}

\clearpage
\appendix
\renewcommand{\thetable}{\thesection.\arabic{table}}
\renewcommand{\thefigure}{\thesection.\arabic{figure}}
\renewcommand{\theequation}{\thesection.\arabic{equation}}

\begin{center}
    \hrule height 4pt \vskip 0.25in
    {\LARGE\bfseries Supplementary Material \par}
    \vskip 0.29in \hrule height 1pt \vskip 0.09in
    \vskip 1em
\end{center}

\emph{In the following supplementary parts to the main manuscript, we provide dataset descriptions in Appendix \ref{section:app:data}, definitions of the physical observables and evaluation metrics in Appendix \ref{section:app:observablesdef}, discuss implementation details for training and finetuning in Appendix \ref{section:app:impldetails}, and provide supplementary experiments in \cref{section:supplementaryresults}. \Cref{section:app:theory} shows that masking preserves a well-defined filtration under the grid approximation and establishes bottleneck stability with respect to the masks, \cref{sec:runtime} reports runtime and memory of the PH computation, \cref{section:app:ablations} ablates homology degrees, the masking strategy and hyperparameters, \cref{section:app:societalimpact} discusses societal impact, and \cref{section:app:additionalfigures} collects additional figures.}

\section{Data}
\label[appendix]{section:app:data}

\subsection{mdCATH}
\label[appendix]{subsection:app:datamdcath}
\mdcath\ \citep{Mirarchi.2024} provides all-atom MD trajectories for 5,398 CATH domains (50--500 residues, mean 137), each simulated in five replicas at five temperatures (320, 348, 379, 413, 450~K) for mainly 500~ns with coordinates recorded at 1~ns frame interval. We adopt the split of \citet{Kapusniak.2026}: an 80/10/10 partition of domains into train, validation, and test, with test domains sharing no more than 20\%
sequence similarity with any training domain (MMseqs2
\citep{Steinegger.2017}, maximum sensitivity). We additionally exclude 48 domains with faulty trajectories
as documented in \url{https://github.com/compsciencelab/mdCATH/issues/7}.
We train on the 450~K
replicas, where (partial) unfolding produces conformational
diversity within the simulated window; lower-temperature replicas remain close to the native basin.

\subsection{Fast Folders}
\label[appendix]{subsection:app:datafastfolders}

While \mdcath\ covers a large structural space, its individual trajectory lengths
on the order of 500 ns are too short to resolve asymptotic behavior. We therefore,
additionally and as \citet{Costa.2025}, evaluate our model on trajectories from \cite{Majewski.2023} who, 
through adaptive sampling, achieved thorough exploration of phase space.
The twelve proteins are the fast folders of \citet{Lindorff.2011}, small domains with $\alpha$-helical, $\beta$-sheet and mixed folds that fold and unfold reversibly on timescales accessible to simulation. 
Data is downloaded as described at 
\url{https://github.com/torchmd/torchmd-protein-thermodynamics/tree/main/Datasets}.
As seen in \cref{tab:app:fastfolderscorpus}, we match Table 1 of \cite{Majewski.2023} for 9 of the 12 proteins. 

For reproducibility, we assemble each per-protein data from the 
\texttt{*filtered.xtc} trajectories: the curated subset already prepared for downstream analysis
paired with their \texttt{filtered.pdb} topologies. 
We excluded part of the protein G trajectories (\texttt{proteing\_3\_trajectories}), a sub-campaign
still containing solvent; For NTL9 and $\lambda$-repressor conservatively using only trajectories which contained the \texttt{``filtered''} suffix in the file name, does not match the aggregated time of \cite{Majewski.2023}. Nonetheless, we estimate the data as more than sufficient as its aggregated time is on the same order of magnitude than those of the remaining 9 proteins ($\sim$0.2-1.4 ms).
For Villin, norleucine residues are mapped to methionine, the closest standard residue, as \marsfm\ supports only standard amino acids.

Following \cite{Majewski.2023}, the first 10\% of frames of all trajectories are discarded to
remove bias from the starting sample, the remainder is featurized by $C_\alpha$-$C_\alpha$ distances
and projected via TICA ($\tau=10$ ns) onto 4 dimensions. 
In this low-dimensional space, per-protein trajectories are clustered into 32 states
via $k$-means++ as in \cite{Costa.2025}.
The native state is identified as the cluster with the minimal centroid distance to the
respective crystal structure. For every protein, we fit an MSM ($\tau=10$ ns) on 
all of its trajectory data (TIC space) to obtain
the protein's stationary distribution with which we weigh the MD frames in our evaluation metrics.

\begin{table}[h]
\centering
\small
\caption{Fast folder data. For each protein, we report
the sequence length, the aggregated simulation time, and the minimum
RMSD to the experimental crystal structure observed across all used
trajectories of \cite{Majewski.2023}.\label{tab:app:fastfolderscorpus}}
\vskip0.5ex
\begin{tabular}{lccc}
\toprule
\textbf{Protein} & Number of residues  & Aggregated time ($\mu$s) & Min.\ RMSD (\AA) \\
\midrule
Chignolin           & 10 &  186.9 & 0.15 \\
Trp-Cage            & 20 &  196.9 & 0.45 \\
BBA                 & 28 &  364.5 & 1.13 \\
WW-Domain           & 34 & 1374.6 & 0.73 \\
Villin              & 35 &  234.7 & 0.47 \\
NTL9                & 39 &  512.9 & 0.32 \\
BBL                 & 47 &  676.5 & 1.55 \\
Protein B           & 47 &  609.4 & 1.21 \\
Homeodomain         & 54 &  199.1 & 0.63 \\
Protein G           & 56 & 1161.7 & 0.55 \\
$\alpha$3D          & 73 &  768.0 & 1.81 \\
$\lambda$-repressor & 80 &  784.0 & 1.52 \\
\bottomrule
\end{tabular}
\end{table}

\section{Definition of physical observables and evaluation metrics}
\label[appendix]{section:app:observablesdef}

\paragraph{RMSD} The root mean-squared deviation (RMSD) is typically computed with respect to a reference
conformation, e.g., a native crystal structure as in \cref{tab:rmsd_fnc_gdt}. 
In case of the pairwise RMSD, we calculate it between a random subset of all pairs of structures 
(once for generated, once for reference MD trajectory)
and compute the Pearson correlation coefficient between their distributions.

\paragraph{RMSF} The root mean-squared fluctuation (RMSF) quantifies atomic flexibility and computes,  
for every $C_{\alpha}$ atom, the deviation of its position along a trajectory. That is for residue $l$, 
\begin{equation}
    \text{\texttt{RMSF}}_l = \sqrt{
        \frac{1}{T} \sum_{t=1}^{T} \Vert \mathbf{r}_{\alpha, l} - \bar{\mathbf{r}}_{\alpha, l} \Vert^2_2
        }
\end{equation}
with $\mathbf{r}_{\alpha, l}$ the coordinate of residue $l$'s $\alpha$-carbon and 
$\bar{\mathbf{r}}_{\alpha, l}$ its mean over time $T$. The RMSF is related to the $\mathrm{B}$ factor,
\begin{equation}
    \mathrm{B}_l = \frac{8 \pi^2}{3} \text{\texttt{RMSF}}_l^2\enspace,
\end{equation}
which can be measured experimentally through X-ray crystallography.

For the RMSF, we compute per-target and global Pearson correlations. 
In the per-target evaluation, the RMSF for a specific protein is compared between reference
MD and generated trajectory. This metric discovers if generated and reference molecules differ in the relation of fluctuation  \emph{between} residues, e.g., if both reference and generated proteins have floppy ends and comparatively rigid interiors.
It would not, however, capture if the generative model prescribes a greater mobility
(or rigidity) to \emph{all} residues. The \emph{global} correlation will pick up exactly
such trends by comparing the distribution of all reference protein simulations with the 
distribution of all generated proteins.

\paragraph{RMSZ} To assess the backbone covalent geometry of generated
structures, we compute root mean-square Z-scores (RMSZ) of bond lengths and
bond angles against the empirical crystallographic targets of
\citet{Engh.1991}. For an observed backbone bond length $\ell_k$ and bond
angle $\theta_k$, with target means $\ell_k^\ast$, $\theta_k^\ast$ and
standard deviations $\sigma_k^{\ell\ast}$, $\sigma_k^{\theta\ast}$,
\begin{equation}
    z_k^{\ell} \;=\; \frac{\ell_k - \ell_k^\ast}{\sigma_k^{\ell\ast}},
    \qquad
    z_k^{\theta} \;=\; \frac{\theta_k - \theta_k^\ast}{\sigma_k^{\theta\ast}},
\end{equation}
and we report
$\mathrm{RMSZ}_{\gamma} = \sqrt{\tfrac{1}{N_{\gamma}}\sum_{k=1}^{N_{\gamma}} (z_k^{\gamma})^2}$
aggregated separately over all backbone bonds ($\gamma=\ell$) and bond
angles ($\gamma=\theta$). Values close to those of the reference MD
ensemble indicate physically realistic covalent geometry, whereas large
RMSZ exposes systematic bond or angle distortions in generated samples.

We obtain crystallographic reference values \texttt{stereo\_chemical\_props.txt}
from AlphaFold's GitHub repository
(see \url{https://github.com/google-deepmind/alphafold/issues/32} in case of retrieval issues).

\paragraph{Radius of Gyration} The radius of gyration (RG) is defined as the root mean-squared distance to the protein's center of mass, i.e.,
\begin{equation}
    \text{\texttt{RG}} = \sqrt{\frac{\sum_{i=1}^N  m_i \Vert \mathbf{r}_i - \bar{\mathbf{r}}\Vert^2_2}{\sum_{i=1}^{N}m_i}}\enspace.
\end{equation}

\paragraph{SSF} We compute the secondary structure fraction (SSF) as the fraction of residues that are part of
\begin{itemize}
    \item $\mathrm{G}$: 3-turn helix
    \item $\mathrm{H}$: 4-turn helix ($\alpha$ helix)
    \item $\mathrm{I}$: 5-turn helix ($\pi$ helix)
    \item $\mathrm{E}$: Extended strand in parallel and/or anti-parallel beta-sheet conformation 
    \item $\mathrm{B}$: Residue in isolated beta-bridge 
\end{itemize}
such that for a protein with $L$ residues,
\begin{equation}
    \text{\texttt{SSF}} = \frac{1}{L} \sum_{l}^{L} \mathbb{I}[\mathrm{DSSP}(l) \in \{\mathrm{G, H, I, E, B}\}]
\end{equation}
with the indicator function $\mathbb{I}$, and the $\mathrm{DSSP}$ (dictionary of protein secondary structure)
denoting the standard algorithm for secondary structure assignment.

\paragraph{Distribution comparison} For \rg\ and \ssf, we compute the value of every frame of the reference MD (all replicas pooled) and of the generated trajectory, and histogram both with 100 bins over their joint range. \texttt{RG KL} and \texttt{SSF KL} report the Kullback--Leibler divergence $D_{\mathrm{KL}}(p_{\mathrm{MD}} \,\|\, p_{\mathrm{gen}})$, with empty bins of $p_{\mathrm{gen}}$ set to $10^{-5}$ before renormalization, and \texttt{SSF JSD} the Jensen--Shannon divergence; both in nats.

\paragraph{Stationary distribution $\pi$} The stationary distribution
$\pi$ is the long-time equilibrium population over the MSM states: the
unique probability vector left-invariant under the transition matrix
$T(\tau)$
\begin{equation}
    \pi^\top T(\tau) = \pi^\top,
    \qquad \sum_i \pi_i = 1,
\end{equation}
i.e., the left eigenvector of $T(\tau)$ at eigenvalue $\lambda_1 = 1$,
independent of the lag time $\tau$. In practice, we obtain this quantity via 
\texttt{deeptime}'s \citep{Hoffmann.2021} \texttt{stationary\_distribution} method on the computed MSM.

\paragraph{FNC} We compute the fraction of native contacts (FNC) with the protocol of \cite{Lewis.2025},
based on \cite{Best.2013}: For all C$_\alpha$ pairs ($C_{\alpha,i},C_{\alpha,j} $) with 
sequence distance $\vert l_i - l_j \vert > 3$ and distance 
$r_{ij} = \vert \mathbf{r}_{i} - \mathbf{r}_{j} \vert < 10 $\AA, the FNC for structure $x$ is
\begin{equation}
   \texttt{FNC}(x) = \frac{1}{P} \sum_{(i, j)} \frac{1}{1 + \exp{(\beta (r_{ij}(x) - \lambda r^0_{ij}))}}
   \label{eq:app:fnc_def}
\end{equation}
with $r_{ij}(x)$ the distance between $i$ and $j$ in configuration $x$, 
$r^0_{ij}$ the corresponding distance in the native state, $\beta= 5 \text{\AA}^{-1}$ 
a smoothing parameter and $\lambda=1.2$ accounting for fluctuations.
Note that this metric corresponds to a coarse-grained version of the FNC, whereas the
all-atom one in \cite{Best.2013} uses cutoff distance $4.5$\AA\ and $\lambda=1.8$.

\paragraph{GDT--TS} The Global Distance Test--Total Score (GDT--TS) \citep{Zemla.2003}
quantifies C$_\alpha$-level similarity to a reference structure across a
range of length scales $\{1,2,4,8\}$ \AA. 
For a conformation $x$ with C$_\alpha$ coordinates $\{\mathbf{r}_i(x)\}_{i=1}^{L}$, rigidly
superimposed onto a reference $\{\mathbf{r}_i^\ast\}_{i=1}^{L}$,
\begin{equation}
    \texttt{GDT-TS}(x)
    \;=\; \frac{1}{4}\sum_{d \in \mathcal{D}}
    \frac{1}{L}\,\bigl|\bigl\{\, i : \|\mathbf{r}_i(x)-\mathbf{r}_i^\ast\| \le d \,\bigr\}\bigr|,
    \qquad \mathcal{D}=\{1,2,4,8\}\,\text{\AA},
\end{equation}
yielding a value in $[0,1]$ that grows toward $1$ as $x$ approaches the
reference.

\paragraph{Folding Free Energy} For \mdcath, we again follow \cite{Kapusniak.2026}
and classify each frame as folded or unfolded using a single threshold
on the \texttt{FNC} defined in 
\cref{eq:app:fnc_def}. A threshold $\texttt{FNC}_{1/2}$ is determined once
from the 320~K reference MD trajectory: we fit a kernel density estimate 
to the distribution of $\texttt{FNC}$ values and take $\texttt{FNC}_{1/2}$ 
to be the location of the deepest minimum within $[0.45, 0.90]$. 
If the KDE has no minimum in this range, 
we fall back to $\texttt{FNC}_{1/2} = 0.70$. 
The same $\texttt{FNC}_{1/2}$ is then applied unchanged to the 450~K MD trajectory and 
to all generative trajectories for that protein, so that all ensembles are compared 
against an identical folded/unfolded boundary.

Rather than a hard cutoff, we use a sigmoid soft assignment to obtain the probability of foldedness of each frame as
\begin{equation}
    p_\text{fold} = \bigl(1 + \exp\{-2s\,[\texttt{FNC} - \texttt{FNC}_{1/2}]\}\bigr)^{-1}
    \label{eq:pfold}
\end{equation}
and report its trajectory average $\bar{p}_\text{fold}$. 
The steepness parameter $s$ controls how sharply the assignment transitions across $\texttt{FNC}_{1/2}$: 
we use $s = 10$, as in \cite{Kapusniak.2026}.
The folding free energy at temperature $T$ then follows from the two-state population ratio:
\begin{equation}
    \Delta G = -k_B T \,\ln\!\left(\frac{\bar{p}_\text{fold}}{1 - \bar{p}_\text{fold}}\right).
    \label{eq:dG}
\end{equation}
For the fast folders, we do have sufficiently more data per protein, such that
we define the probability to be folded as the stationary distribution value 
of the native cluster,  $\bar{p}_\text{fold} = \pi_{\text{native}}$. Otherwise,
the evaluation in \cref{eq:dG} remains the same.
We obtain the underlying stationary distributions by fitting an MSM (lag time 10 ns) once on all available MD trajectories, and on all generated trajectories.

\paragraph{CK}\label{par:app:ck}
The Chapman--Kolmogorov (CK) property requires that the transition matrix of a Markovian system satisfies
\begin{equation}
    T(\tau)^k = T(k\tau).
\end{equation}
Following \cite{Prinz.2011}, we test this by coarse-graining a 10-state MSM into 3 macrostates via PCCA+ \citep{Roblitz.2013} and,
for each macrostate $A$, 
comparing the probability to remain in the macrostate, $A \to A$, at lag $k\tau$ 
predicted by $T(\tau)^k$ with the empirical estimate from MD counts.

For convenience, we restate important equations from \cite{Prinz.2011}. For a Markov Model
transition matrix $T(\tau)$, let $\boldsymbol{\pi}$ be its stationary distribution.
Its restriction to macro state $A$ is defined as
\begin{equation}
    w_i^{A} := 
        \begin{cases} 
            \frac{\pi_i}{\sum_{j \in A} \pi_j} \ & i \in A  \\
            0 \ & i \notin A \\
        \end{cases}.
\end{equation}
The probability to remain in $A$ in time step $k \tau$ estimated from the reference MD trajectory reads
\begin{equation}
    p_{MD}(A, A; k \tau) = \sum_{i \in A} w_i^A p_{MD}(i, A; k \tau)
\end{equation}
where the estimated probability to transition from frame $i$ into state $A$ is
\begin{equation}
    p_{MD}(i, A; k \tau) 
        = \frac{\sum_{j \in A} c_{ij}^{obs}(k \tau)}{\sum_{j=1}^n c_{ij}^{obs}(k \tau)}
        \label{eq:app:ck_p_md}
\end{equation}
with transition counts $c_{ij}^{obs}(k \tau)$ from state $i$ to state $j$ in time $k \tau$. 
The corresponding probability according to the MSM would be 
\begin{equation}
    p_{MSM}(A, A; k \tau) = \sum_{i \in A} (w_i^A)^\top T^k(\tau)
\end{equation}

For the actual computation of the \texttt{RMSE} in \cref{tab:msm-quality}, we 
compute the root mean-squared error of the probability estimate per replica, averaged over 
all macro states $A \in \{ 1, 2, 3\}$ and $k \in \{1, 2\}$
\begin{equation}
    \texttt{CK RMSE} =  \sqrt{
    \frac{1}{\text{\# macrostates \#k}} \sum_{A, k} (p_{MD}(A, A; k\tau) - p_{MSM}(A, A; k\tau))^2
}
\end{equation}
For that domain, the result is then averaged over all 5 replicas.

As can be seen from \cref{eq:app:ck_p_md}, estimating the required probability
to remain in state $A$ within time $k \tau$, $p_{MD}(A, A; k\tau$), 
from a single held-out replica yields noisy estimates for small $c_{ij}^{obs}$ 
that become ill-defined whenever $c_{ij}^{obs}$ becomes 0. That is, if
$A$ is not visited frequently or at all within the first $T - k\tau$ frames,
with $T$ the length of the replica trajectory. This can happen easily for a single held-out 
short 500 frame trajectory.

\section{Implementation details}
\label[appendix]{section:app:impldetails}

\subsection{Vectorizing persistence diagrams}
\label[appendix]{app:sec:vectorization}
In order to use persistence diagrams as inputs to standard machine learning models one first needs to represent them as vectors. To this end, we employ the vectorization method from \cite{Hofer.2019} based on exponential structure elements. Given a (center) location parameter $\mu\in \mathbb R^2$ and a sharpness parameter $\sigma \in \mathbb R^2$, \cite{Hofer.2019} define 
\begin{equation}
    t_{\mu,\sigma}: \mathbb R^2 \to \mathbb R\enspace, 
     \quad x \mapsto \exp\left(- \sum_{i\in \{1,2\}}\sigma_i^2 (\mu_i - x_i)^2\right)
\end{equation}
This map is applied to persistence diagrams, or rather all points $(b,d)$ in a persistence diagram with finite death-time $d<\infty$, i.e., a multiset of points in the plane $\mathbb R^2$ with $d>b$, after transforming them to (birth,lifetime)-coordinates $(b,d-b)$ and the coordinate stretch 
\begin{equation}
    \tau_\nu: \mathbb R \times (0,\infty) \to \mathbb R^2\enspace, \quad (b,d-b) \mapsto \begin{cases}
        (b,d-b) & d-b \ge \nu \\ (b, 2\nu - \tfrac{\nu^2}{d-b}) & d-b < \nu
    \end{cases}
    \enspace.
\end{equation}
The map $(b,d) \mapsto t_{\mu,\sigma} \circ \tau_\nu (b, d-b)$ is then called an \emph{exponential structure element}, and as $\tau_\nu$ stretches lifetime coordinates lower than $\nu$ towards $-\infty$, its composition with $t_{\mu,\sigma}$ maps those points close to zero.
Summation over all points in the persistence diagram produces a map to $\mathbb R$ via
\begin{equation}
    \dgm \mapsto \sum_{(b,d) \in \dgm} t_{\mu,\sigma} \circ \tau_\nu (b, d-b) \enspace,
\end{equation}
and by construction, the contribution of points tends to zero as the lifetime decreases to zero.
Repeating this for multiple centers $\mu_1,\dots, \mu_d$ and sharpness parameters $\sigma_1,\dots, \sigma_d$, i.e., using $d$ different structure elements, and concatenating the output then gives a vectorization map to $\mathbb R^d$ parametrized by $2d + 2d + 1 = 4d+1$ parameters.

While in principle, these parameters can be learned within a neural network, see \cite{Hofer.2019}, we directly fit them to the data so that the vectorized persistence diagrams are fixed features for learning. To do so, we use a strategy as in ATOL \citep{Royer.2021}.
We first stack all persistence diagrams in the training split of a dataset, producing a multiset of points in $\mathbb R^2$ of large cardinality and apply the (birth,lifetime)-transform. We then fit the centers $\mu_j$ using $k$-means with $k$-means++ initialization \citep{Arthur.2007}. As the respective sharpness parameter $\sigma_j$ we use $\sigma_j = (\tfrac 2 \delta_j, \tfrac 2 \delta_j)$ with $\delta_j = \min_{k\neq j} \|\mu_j -\mu_k\|_2$. 
This leaves two hyperparameters that need to be selected: the number $d$ of centers, and the quantile $q$ (on the training split) that is used to determine the stretch parameter $\nu$.
In all applications we set $d=128$ and $q=0.01$.

\subsection{Setup for Representation Quality Probes}
\label[appendix]{app:sec:tasks_setup}
The inputs to the probing models are vectorized persistence diagrams of dimension 384, i.e., 128
dimensions per $H_0$, $H_1$ and $H_2$ diagram, computed with the stretch quantile $q = 0.01$ for every
simplicial complex (\cref{app:sec:vectorization}). Each probe is either a linear map or an MLP with ELU
activations and hidden dimensions (128, 64). For CATH classification the probe maps to four outputs; for
observable regression it maps to a scalar, and one probe is trained per observable. Features are
standardized with statistics of the training split. The VAMPNet uses an MLP with hidden dimensions
(256, 128, 64, 32, 16) as backbone and learns 7-dimensional kinetic embeddings. For its domain-wise loss,
a minibatch is constructed from 4 domains, 4 replicas per domain, and 400 time-lagged pairs (all sampled
uniformly at random).

\paragraph{CATH classification}
The frame vectors of each replica are averaged over time, so that each replica contributes one sample
(about 21k / 2.7k / 2.4k samples for training / validation / testing). Probes minimize a class-balanced
cross-entropy with batch size 128 for at most 100 epochs, with early stopping on the validation balanced
accuracy (patience 20) and restoration of the best weights. We report the balanced accuracy on the test
split, averaged over three seeds.

\paragraph{Observable regression}
Every frame of all five 450~K replicas is used, i.e., about 9.9M / 1.2M / 1.1M frames for
training / validation / testing. Probes minimize the mean squared error on the observable in its physical
units, with batch size 16,384. Training runs for at most 200 epochs and stops early once the validation
$R^2$ has not improved by at least $10^{-3}$ for 20 consecutive epochs; the weights of the epoch with the
best validation $R^2$ are restored. All runs stopped early, after 21--67 epochs. We report the test $R^2$
over all test frames, averaged over three seeds.

The setup for the (($\phi$, $\psi$) + C$_\alpha$)-baseline is as follows. For each MD simulation of a domain, the 7 per-residue features, i.e., angles and C$_\alpha$ coordinates are stacked along the residues and time steps, resulting in a tensor of shape $(\text{\#timesteps}, 7,  \text{\#residues})$. These are then input to a 1D convolutional network, where the time steps serve as a batch and the 7 per-residues features as input channels; the convolution kernel operates along the residue axis.
The convolutional network consists of three 1D convolutional layers with kernel size 7 and stride 1, ReLU activations, and output channels of size 16, 64 and 256.
This is followed by an average pooling layer operating along the residue axis, producing embeddings of shape $(\text{\#timesteps}, 256)$ and a final linear layer projecting the embeddings either to $\mathbb R$ for regression or to $\mathbb R^4$ for classification. Gradients are accumulated over multiple domains to mimic batching.

\paragraph{Optimization} All probes are trained using ADAM with a cosine annealing schedule stepped
per epoch, an initial learning rate of $10^{-3}$ and weight decay $10^{-3}$. The VAMPNet is trained
using ADAM with an initial learning rate of $5 \cdot 10^{-4}$ and weight decay $10^{-3}$ for at most
200 epochs of 50 minibatches. Its learning rate is halved whenever the \texttt{vamp-2} score on the
validation domains has not improved by a relative $10^{-3}$ for 5 epochs, down to $10^{-6}$; training
stops after 5 epochs without improvement, and the weights of the best epoch are restored. Early
stopping uses the \texttt{vamp-2} score of the continuous embedding, not the MSM measures we report.
Gradients are clipped at a global norm of 5, and the loss adds an $L_2$ penalty of $10^{-3}$ times
the mean squared embedding.
The end-to-end trained graph encoders deviate from this as described in \cref{app:sec:mlbaselines}.

\subsection{ML Baselines}
\label[appendix]{app:sec:mlbaselines}

We compare the persistence-based representations against two kinds of learned baselines.
\emph{Learned end-to-end baseline models} (\texttt{EGNN}, \texttt{GotenNet}) are trained jointly with the probe head on each task. 
\emph{Learned representations baselines} (\texttt{ESM2}, \texttt{ESM-IF1} and their fusion) provide a fixed vector per
domain or per frame, computed once, on top of which only a probe head is trained, exactly as for the
vectorized persistence diagrams. 

\subsubsection{Learned end-to-end baseline models}

\paragraph{Input graphs}
Both graph encoders operate on the same residue-level graph, with one node per residue, positioned at its C$_\alpha$ atom.
As node features the residue identity is mapped to 20 classes. Edges are determined by implementing
a radius graph with cutoff 1.0~nm (10~\AA) and at most 32 neighbors per node,
without self-loops. The original models were designed for small molecules
(\texttt{EGNN} on fully connected QM9 graphs), which is infeasible for domains of 50--500 residues.

For CATH classification each domain contributes the first frame of each of its five replicas
($\sim$ 21k training, 2.6k validation and 2.4 k testing frames).
For observable regression we use every 50th frame of every replica ($\sim$ 199k training, 25k
validation and 22.7k testing frames); end-to-end training on every frame was outside of our compute budget.

\paragraph{EGNN}
We use the reference implementation of \cite{Satorras.2021} with the configuration of the
original QM9 experiments: 7 layers, hidden and output width 128, attention and residual connections enabled,
no coordinate normalization and no $\tanh$ on the
coordinate update. The node-input is the one-hot residue identity; the only edge input is the squared
distance. 
We adapt the model to the larger sized graphs of \mdcath\ domains in three ways:
the radius graph above replaces the fully connected graph;
connectivity is computed once from the input coordinates and not rebuilt as the layers
update the coordinates; and the updated coordinates are discarded, only the invariant node features
$h_i$ being pooled into one vector per graph.

\paragraph{GotenNet}
We use \texttt{GotenNetWrapper} from the authors' package \citep{Aykent.2025} (version 1.1.3), with 
scalar width 128, 8 interaction layers, $l_{\max} = 1$, 32
radial basis functions, a cosine cutoff at 1.0~nm and at most 32 neighbors per node. Residue identities
enter through \texttt{GotenNet}'s atom-type embedding.
The radius graph is built inside \texttt{GotenNet} from the C$_\alpha$ coordinates, and the
scalar node features after the last layer are pooled.

\paragraph{Training of the graph encoders}
The encoders are mean-pooled over nodes and read out by the same two-layer MLP head (128, 64; ELU).
Graphs are packed into batches of at most 8,000 nodes to bound memory across domain sizes.
Otherwise, training follows the probes (\cref{app:sec:tasks_setup}), with three deviations:
a weight decay of $10^{-4}$ instead of $10^{-3}$ (at $10^{-3}$ both encoders stall, e.g., at a validation $R^2$
of 0.22--0.52 on \rg, whereas $10^{-4}$ reaches 0.83 after one epoch), gradient clipping at a global
norm of 1.0 as \texttt{EGNN} would otherwise diverge, and a budget of at most 200 epochs also for classification.

On one NVIDIA H100 GPU, one regression epoch takes about 7.5~min for \texttt{GotenNet} and 1.8~min for
\texttt{EGNN}.

\subsubsection{Learned Representations}
\paragraph{ESM2}
We use the pretrained \texttt{ESM2} models \citep{Lin.2023} \texttt{esm2\_t6\_8M\_UR50D} (6 layers,
320 dimensions), \texttt{esm2\_t12\_35M\_UR50D} (12 layers, 480 dimensions), \texttt{esm2\_t30\_150M\_UR50D} (30 layers, 640 dimensions) and
\texttt{esm2\_t33\_650M\_UR50D} (33 layers, 1,280 dimensions). Each domain's residue sequence is tokenized
with the model's alphabet; the final-layer residue representations are averaged over residues,
yielding one vector per domain. As the
sequence is identical in every frame, \texttt{ESM2} enters CATH classification as a per-domain
 rather than a per-replica vector and observable regression only through the fusion model below.

\paragraph{ESM-IF1}
We use the encoder of the inverse folding model \texttt{esm\_if1\_gvp4\_t16\_142M\_UR50}
\citep{Hsu.2022}. Its input is the backbone (N, C$_\alpha$, C) of a single
frame with its 512-dimensional per-residue outputs averaged over residues to give one vector per frame. 
For CATH classification, as for PH vectorizations, each replica is represented by the 
trajectory-averaged vector.

\paragraph{Fusion (\texttt{ESM2\_35m}, \texttt{ESM-IF1})}
The domain's \texttt{ESM2\_35m} vector (480 dimensions) is repeated for every frame and concatenated with
that frame's \texttt{ESM-IF1} vector (512 dimensions), giving 992 dimensions of which only the
\texttt{ESM-IF1} part varies within the trajectory. \texttt{ESM2\_35m} was chosen as the sequence representation since it is performing best in the ESM2 family on CATH classification (see \cref{table:app:clf}).

\subsection{ESM Train-Test overlap}
\label{sec:app:esmoverlap}
    The ESM2 family was trained on $\sim$65 million unique sequences from UniRef50 \citep{Suzek.2015},  v2021\_04,
    which makes an overlap with our test set likely but cannot be verified as training IDs are
    not published.

ESM-IF1 (\texttt{esm\_if1\_gvp4\_t16\_142M\_UR50}) was pretrained on $\sim$16k structures from CATH v4.3 and 12 million AlphaFold2 predicted ones from UniRef50 v2021\_04. An overlap with the latter 12 million structures is possible, but can again not be verified as these training IDs are not public.
    The PDB ids of the CATH portion can be retrieved at \url{https://dl.fbaipublicfiles.com/fair-esm/data/cath4.3_topologysplit_202206/splits.json}.
    43.9\% (214) of our 488 test domains occur exactly in ESM-IF1's CATH training set with 207
    as an exact match and 7 as re-depositions under another PDB id. In every case,
    the \mdcath\ sequence is an exact segment of the ESM-IF1 training chain. For these, crystal-structure
    alignment confirms structural identity (median qTM = 1.0). Of the 274 remaining test domains,
    a further 47 reach qTM $\geq$ 0.9 against an ESM-IF1 training structure, versus 3 
    when the same domains are measured against our own training split. In total, 53.5\% of the test set
    is either identical or near-identical to something ESM-IF1 trained on. For context, our own
    train/test split has no detectable sequence homology (0/488 domains at $\geq$20\% identity, as \mdcath{}
    is drawn from CATH's S20 non-redundant set). This entails substantial pretraining--evaluation
    overlap. We report the ESM-IF1 and (ESM2, ESM-IF1) results for completeness, but they should be digested
    with this caveat in mind.%

\subsection{VAMPNet}
\label[appendix]{subsection:app:vampnet}
Formally, given coordinate trajectories $\{\mathbf{x}_t\}_t$ (e.g., vectorized persistence diagrams) of different protein domains $d$ drawn from some distribution $p$, we want to optimize $f_{\boldsymbol{\theta}}$ such that the induced latent trajectories $\{f_{\boldsymbol{\theta}}(\mathbf{x}_t)\}_t$ maximize the expected \texttt{vamp-2} score (for some fixed lag time $\tau$). That is, we search for 
\begin{equation}
\label{eqn:objectivegeneral}
    \begin{split}
    \boldsymbol{\theta}^\star =
    \arg\max_{\boldsymbol{\theta}} \mathop{\mathbb{E}}_{d \sim p} \underbrace{\left[
    \left\|
        \left(\mathbf{C}_{00}^{(d)}\right)^{-\nicefrac{1}{2}}
        \mathbf{C}_{01}^{(d)}
        \left(\mathbf{C}_{11}^{(d)}\right)^{-\nicefrac{1}{2}}
    \right\|_F^2 \right]}_{\text{\texttt{vamp-2} score}}
    \enspace,
    \end{split}
\end{equation}
where $\mathbf{C}_{00}^{(d)}$,$\mathbf{C}_{01}^{(d)}$ and $\mathbf{C}_{11}^{(d)}$ denote the domain-specific covariance matrices constructed from different combinations of time-lagged pairs $(\mathbf{z}_t, \mathbf{z}_{t+\tau}) = (f_{\boldsymbol{\theta}}(\mathbf{x}_t), f_{\boldsymbol{\theta}}(\mathbf{x}_{t+\tau}))$ of VAMPNet representations.

In practice, we rely on stochastic optimization of an empirical estimate of \cref{eqn:objectivegeneral} based on multiple simulation replicas $\mathcal{R}_d$ of each domain $d \in \mathcal{D}$. Specifically, we first sample a mini-batch of domains $\mathcal{D}_{b} \subset \mathcal{D}$; then, for each sampled domain $d \in \mathcal{D}_b$, we select $N_R$ replicas $\mathcal{R}_{d,b} \subseteq \mathcal{R}_d$ at random and, within each replica, a set of $N_T$ random time points $\mathcal{T}_{d,r}$ (without replacement). Upon collecting all time-lagged pairs and letting 
$\mathbf{z}_{r,t} = f_{\boldsymbol{\theta}}(\mathbf{x}_{r,t})$ denote the VAMPNet representation of $\mathbf{x}_{r,t}$, we construct the empirical per-domain covariance matrices
\begin{equation*}
    \begin{split}
        \widehat{\mathbf{C}}_{00}^{(d)} = \frac{1}{N_d}\sum_{\substack{r \in \mathcal{R}_{d,b} \\ t \in \mathcal{T}_{d,r}}} \mathbf{z}_{r,t} \mathbf{z}_{r,t}^\top\enspace,
        \widehat{\mathbf{C}}_{01}^{(d)} = \frac{1}{N_d}\sum_{\substack{r \in \mathcal{R}_{d,b} \\ t \in \mathcal{T}_{d,r}}} \mathbf{z}_{r,t} \mathbf{z}_{r,t+\tau}^\top\enspace,
        \widehat{\mathbf{C}}_{11}^{(d)} = \frac{1}{N_d}\sum_{\substack{r \in \mathcal{R}_{d,b} \\ t \in \mathcal{T}_{d,r}}} \mathbf{z}_{r,t+\tau} \mathbf{z}_{r,t+\tau}^\top\enspace,
    \end{split}
\end{equation*}
with $N_{d} := \sum_r |\mathcal{T}_{d,r}|$ and compute the empirical \texttt{vamp-2} loss accordingly.
 Finally, using the trained network $f_{\boldsymbol{\theta}}$, we map all trajectory frames to the latent space and, for each domain, discretize the representations of each replica into $k$ states using $k$-means++. The resulting discrete trajectories are then used to estimate MSM transition matrices via maximum likelihood at lag time $\tau$.

\subsection{MarS-FM}

\paragraph{Training}
Training follows the setup of \cite{Kapusniak.2026} exactly. We use a transformer 
comprising a stack of Invariant Point Attention layers \citep{Jumper.2021} (4 heads, 32-dimensional head size, 8 query-key and 8 value points) 
followed by a stack of 5 modified DiT blocks \citep{Peebles.2023, Jing.2024} with 384-dimensional token embeddings 
and 16-head multi-head attention.

Optimization uses ADAM with a constant learning rate of $10^{-4}$ and an exponential moving average with decay 0.999. Training pairs are drawn from the per-protein MSM with 2 clusters and 12 sequences per cluster. We train for 1,000 epochs at batch size 8. For generation, the learned velocity field is integrated with the adaptive \texttt{dopri5} solver 
from \texttt{torchdiffeq} \citep{Chen.2018} using absolute tolerance $10^{-6}$, relative tolerance $10^{-3}$ evaluated at 50 equally spaced time points.

\paragraph{Generation}

We follow the hierarchical tree-sampling scheme of \cite{Kapusniak.2026}.
Starting from the initial frame, we first generate 200 conformations in parallel. In a second pass, we generate 300 additional conformations, 
each conditioned on one of the first-step samples, 
yielding 500 samples in total. 
The first-layer branching factor of 200 corresponds to the default identified 
as near-optimal in \cite{Kapusniak.2026}.

\subsection{Fast folders}
\label[appendix]{subsection:app:implfastfolders}
\paragraph{Finetuning}
To adapt to the dynamic regime of the fast folders (\cref{subsection:app:datafastfolders})
we finetune our network on the 348 K simulations of \mdcath\
by the following strategy: 
The default \marsfm\ training batch consists of 2 clusters with 12 samples each for 8 protein domains.
One of the two sampled clusters always corresponds to the initial frame cluster, 
usually a comparatively compact conformation, while the second is sampled randomly. 
Thereby, the model obtains a strong learning signal for generating \emph{from} compact
structures toward extended structures. The fast folding dataset, however, 
specifically seeks out folding events.
To match this dynamic regime, we flip the role of the starting frame cluster, sampling clusters 
that transition \emph{into} it.

To adapt to the faster dynamics, we reduce lag time $\tau$ with respect to which MSMs are fitted (for sampling training pairs), to 10 ns.
Furthermore, VAMPNet training on the 348 K dataset (persistence diagram) vectorizations of \mdcath\ is also run at a lag time of 10 ns.

We finetune for 300 epochs using ADAM with a one-cycle schedule (\texttt{OneCycleLR} in PyTorch \citep{Pazke.2019}), i.e., a linear warm-up to a peak learning rate of $10^{-5}$ over the first 10\% of training steps, followed by cosine annealing.
The \marsfm\ baseline is finetuned with the identical protocol and differs only in the features from which its MSMs are estimated.

\paragraph{Generation} For generation, we sample each of the 32 clusters
32 times uniformly for source conformations as input for our generative model, generating 500 samples in a tree sampling manner as prescribed by
\cite{Kapusniak.2026}: We generate 200 conformations in the first step, and 300 more 
in the second with the initial 200 as input. This yields per protein $32 \times 32 \times 500=512{,}000$ generated conformations.

\subsection{Computing infrastructure} 
\label[appendix]{app:compute}
All experiments were run on an Ubuntu Linux system (24.04), running kernel 6.8.0-83-generic,
with two AMD EPYC 7313 16-Core Processors (64 threads), 512 GB of main memory, and two NVIDIA H100 NVL GPUs.

\section{Supplementary results}
\label[appendix]{section:supplementaryresults}

\subsection{Probing Representation Quality}
\label[appendix]{subsection:app:probes} 
In \cref{table:app:clf}, we provide CATH classification results for additional sequence-based models, \texttt{ESM2\_8m}, \texttt{ESM2\_150m} and \texttt{ESM2\_650m}, as well as the fusion model, i.e. concatenated representations \texttt{(ESM2\_35m, ESM-IF1)}.
\begin{table}[h]
\vspace{-0.45cm}
\caption{Additional Per-replica classification of CATH class. Balanced accuracy (mean $\pm$ std over 3 runs). \textbf{Best}/\underline{second-best} PH-based result.\label{table:app:clf}}
\vspace{-0.3em}
\centering{
\begin{small}
\begin{tabular}{lcc}
\toprule
            & Linear & MLP \\ %
\midrule
\mfloodph  & \underline{79.1} \std{0.7}	  & \textbf{81.2} \std{2.1} \\
\floodph             & 78.4 \std{2.0}                & \underline{80.0} \std{1.6} \\
\alphaphca           & 78.4 \std{1.0}	              & 78.2 \std{5.8} \\
\alphaph             & 71.3 \std{3.4}	              & 74.5 \std{3.1} \\
\ripsphca	         & \textbf{79.5} \std{0.8}        & 77.4 \std{5.3} \\
\midrule
(\rg, \ssf, \fnc, \rmsd)    & 54.3 \std{0.6} &  60.5 \std{0.4} \\
($\phi$, $\psi$) + C$_\alpha$ coords.  & -- & 70.8 \std{0.1}  \\
\midrule
\texttt{ESM2\_8m} & 76.9 \std{1.3} & 75.5 \std{2.6} \\
\texttt{ESM2\_35m} & 79.5 \std{0.2} & 77.3 \std{0.6} \\
\texttt{ESM2\_150m} & 75.4 \std{1.5} & 74.0 \std{0.7} \\
\texttt{ESM2\_650m} & 73.5 \std{0.3} & 76.1 \std{1.4} \\
\textcolor{lightgray}{\texttt{ESM-IF1}} & \textcolor{lightgray}{82.7} \std{1.8} & \textcolor{lightgray}{78.7} \std{3.6} \\
\textcolor{lightgray}{\texttt{(ESM2\_35m, ESM-IF1)}} & \textcolor{lightgray}{81.5} \std{1.6} & \textcolor{lightgray}{78.8} \std{1.8} \\
\midrule
\texttt{EGNN} & -- & 71.8 \std{5.1} \\
\texttt{GotenNet} & -- & 75.5 \std{7.9} \\
\bottomrule
\end{tabular}
\end{small}}
\vspace{-0.4cm}
\end{table}

\Cref{table:app:regression} lists the regression results of \cref{table:regression} with standard deviations over three seeds. All models are evaluated on every test frame, except \texttt{EGNN} and \texttt{GotenNet}, which are trained and evaluated on every 50th frame because of their training cost (\cref{app:sec:mlbaselines}).

\begin{table}[h]
\caption{Regression of physical observables per time point across testing domains of \mdcath\ (\cref{table:regression}). Coefficient of determination $R^2$ $\uparrow$ (mean $\pm$ std over 3 seeds). \textbf{Best}/\underline{second-best} PH-based result.\label{table:app:regression}}
\centering{
\begin{small}
\begin{tabular}{lcccc}
\toprule
 & \fnc & \rg & \rmsd & \ssf \\
\midrule
\multicolumn{5}{c}{Linear} \\
\midrule
\mfloodph & \textbf{0.691} \std{.002} & \underline{0.728} \std{.001} & \underline{0.683} \std{.001} & \underline{0.652} \std{.003} \\
\floodph & \underline{0.686} \std{.003} & 0.719 \std{.002} & 0.675 \std{.001} & \textbf{0.658} \std{.004} \\
\alphaph & 0.668 \std{.000} & \textbf{0.831} \std{.001} & \textbf{0.708} \std{.002} & 0.623 \std{.000} \\
\alphaphca & 0.669 \std{.002} & 0.675 \std{.001} & 0.656 \std{.000} & 0.650 \std{.003} \\
\ripsphca & 0.630 \std{.001} & 0.559 \std{.001} & 0.597 \std{.001} & 0.599 \std{.002} \\
\addlinespace
\textcolor{lightgray}{\texttt{ESM-IF1}} & \textcolor{lightgray}{0.918} \std{.001} & \textcolor{lightgray}{0.904} \std{.001} & \textcolor{lightgray}{0.822} \std{.000} & \textcolor{lightgray}{0.933} \std{.001} \\
\textcolor{lightgray}{\texttt{(ESM2\_35m, ESM-IF1)}} & \textcolor{lightgray}{0.929} \std{.001} & \textcolor{lightgray}{0.905} \std{.000} & \textcolor{lightgray}{0.829} \std{.001} & \textcolor{lightgray}{0.932} \std{.000} \\
\midrule
\multicolumn{5}{c}{MLP} \\
\midrule
\mfloodph & \underline{0.860} \std{.002} & 0.861 \std{.000} & 0.781 \std{.001} & 0.842 \std{.000} \\
\floodph & \textbf{0.863} \std{.001} & \underline{0.864} \std{.000} & \underline{0.782} \std{.000} & \underline{0.853} \std{.001} \\
\alphaph & 0.858 \std{.001} & \textbf{0.921} \std{.000} & \textbf{0.798} \std{.000} & 0.815 \std{.000} \\
\alphaphca & 0.859 \std{.001} & 0.849 \std{.001} & 0.772 \std{.000} & \textbf{0.854} \std{.000} \\
\ripsphca & 0.840 \std{.000} & 0.772 \std{.000} & 0.732 \std{.002} & 0.828 \std{.001} \\
\addlinespace
($\phi$, $\psi$) + C$_\alpha$ coords. & 0.762 \std{.015} & 0.332 \std{.018} & 0.578 \std{.003} & 0.841 \std{.003} \\
\addlinespace
\textcolor{lightgray}{\texttt{ESM-IF1}} & \textcolor{lightgray}{0.929} \std{.000} & \textcolor{lightgray}{0.946} \std{.000} & \textcolor{lightgray}{0.849} \std{.001} & \textcolor{lightgray}{0.934} \std{.000} \\
\textcolor{lightgray}{\texttt{(ESM2\_35m, ESM-IF1)}} & \textcolor{lightgray}{0.942} \std{.000} & \textcolor{lightgray}{0.944} \std{.001} & \textcolor{lightgray}{0.853} \std{.005} & \textcolor{lightgray}{0.934} \std{.000} \\
\addlinespace
\texttt{EGNN} & 0.878 \std{.003} & 0.955 \std{.011} & 0.799 \std{.026} & 0.863 \std{.005} \\
\texttt{GotenNet} & 0.879 \std{.004} & 0.953 \std{.025} & 0.808 \std{.016} & 0.868 \std{.002} \\
\bottomrule
\end{tabular}
\end{small}}
\end{table}

\subsection{Dynamics}

\subsubsection{Validity of mdCATH MSM Evaluation}
\label[appendix]{subsubsection:app:mdcath:validmsm} 
When applying VAMP naively to short trajectories, as they are present in \mdcath,
one may run into problems overfitting. This is one reason
for learning a common feature map across domains, increasing and diversifying the available data on which the
embedding is estimated from $\sim\!2.3\mu$s per domain to $\sim\!10$ms using the whole training set.
In general, to deal with finite data we follow \cite{McGibbon.2015}, who argue that in the case of the
cross-validation mean test set generalized matrix Rayleigh quotient (GMRQ) is the proper metric for model selection.
The GMRQ $\mathcal{R}$ is an equivalent formulation of the VAMP score.

Equation 24 in \cite{McGibbon.2015} proposes to calculate the mean over the training folds to
gauge overfitting by the difference between train and test $\mathcal{R}.$ 
\begin{table}[h]
    \caption{Test and train VAMP2. Large differences would indicate overfitting on the training trajectories.
    \label{tab:app:msmoverfit}}
    \centering
    \begin{tabular}{lrr}
        \toprule
        Feature & test VAMP2 & train VAMP2 \\
        \midrule
        \mfloodph & 1.503 \std{0.010} & 1.503 \std{0.001} \\
        (\rg, \ssf) & 1.513 \std{0.004} & 1.559 \std{0.001} \\
        \bottomrule
    \end{tabular}
\end{table}
We compare this difference for \mfloodph\ and the \marsfm\ combination of (\rg, \ssf) in \cref{tab:app:msmoverfit},
showing that \mfloodph\ does not display a gap between train and test data.

It is a major strength of MSMs that the underlying simulations do \emph{not} need
to sample from equilibrium distributions \citep{Nueske.2017, Schuette.1999}.  
\citet{Nueske.2017} consider trajectories which neither start nor end in the
global equilibrium as the simulation time is too short, and state that short trajectories
starting from local equilibria (\mdcath\ equilibrates every replica for 20~ns) suffice, if they
last longer than the system's largest timescale $t_2$. We estimate $t_2$ implied by MSMs
fitted on \mfloodph\ and (\rg, \ssf) features, resp., and report the $t_2$-mean per class
$\pm$ its standard error in \cref{tab:app:t2}.

\begin{table}[h]
    \caption{Largest implied timescale per CATH class. \label{tab:app:t2}}
    \centering
    \begin{tabular}{lrr}
        \toprule
        CATH class & $t_2$(\mfloodph) & $t_2$(\rg, \ssf) \\
        \midrule
        mainly-alpha & 67.2 \std{5.6} & 84.1 \std{9.1} \\
        mainly-beta & 153.8 \std{15.3} & 165.3 \std{14.4} \\
        alpha-beta & 122.6 \std{9.8} & 133.5 \std{9.7} \\
        few sec. struct. & 34.7 \std{4.9} & 50.7 \std{4.9} \\
        \hline
    \end{tabular}
\end{table}

\begin{figure}[ht]
\centering
\includegraphics[width=\textwidth]{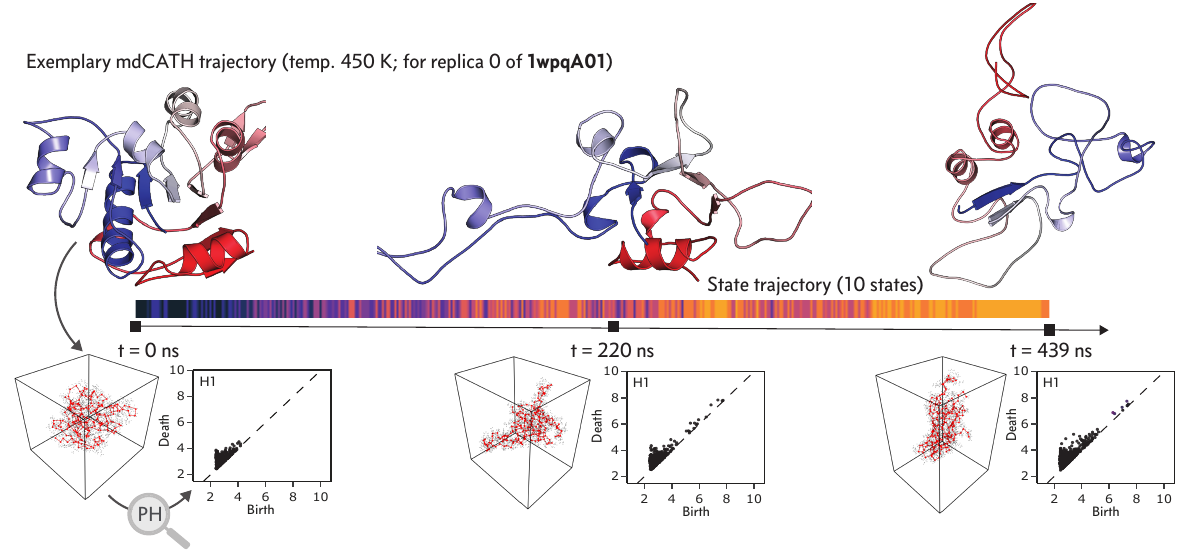}
\caption{Unfolding at 450~K, shown for replica 0 of the \mdcath\ domain \textbf{1wpqA01}. \textbf{Top}: conformations at three time points and the state trajectory obtained from the 10-state discretization of the shared \mfloodph\ VAMPNet embedding. \textbf{Bottom}: point clouds (C$_\alpha$ atoms in red) and $H_1$ persistence diagrams of the masked Flood complex at the same time points.\label{fig:app:unfolding}}
\end{figure}

Notably, the estimates per CATH class follow the trends reported by \cite{Mirarchi.2024} when introducing \mdcath.
For the 450~K simulations, the significant process is the unfolding of the protein domain (see \cref{fig:app:unfolding} for an example).
According to \cite{Mirarchi.2024}, unfolding occurs for $\alpha$ rich structures abruptly within
100~ns, for $\alpha$-$\beta$ and $\beta$ structures somewhat more slowly (cf.~\cite{Mirarchi.2024}, Figure S1).
Thus, according to \citet{Nueske.2017}, \mdcath\ trajectories suffice to estimate valid MSMs.

\subsubsection{MSM Evaluation}
\label[appendix]{subsubsection:app:mdcath:msm}
TICA, as a special linear case of VAMP, allows for dimensionality reduction onto the components of
maximal autocorrelation of a time series. It still needs some input features, and it needs to be applied to (the MD trajectories of) each domain separately. Importantly, it requires inverting the instantaneous covariance matrix; a matrix of size $(\mathrm{num_{features}}, \mathrm{num_{features}})$ and rank $\le \mathrm{num_{observed~timepoints}}$. This renders the use of TICA challenging on \mdcath, where one typically has only 2k observations per domain, and the domains consist of up to 500 residues. In particular, the features used in \cite{Scherer.2019} (for a different kind of data) are ill-suited. The distance-based features are quadratic in $\mathrm{num_{residues}}$, and even the per-residue features ($C_{\alpha}$, torsion), except SASA (which only uses 1 feature per residue), are already so many that TICA overfits, especially on large domains.

In \cref{tab:app:msm} we report MSM quality metrics on the validation domains based on TICA embeddings of three features: $C_\alpha$ coordinates, Torsion (sine-cosine pairs of all backbone and sidechain angles) and $(\phi,\psi) + C_\alpha$ (again sine-cosine pairs of the angles). In correspondence with our VAMPNet implementation, we project to the 7 first TICA components. For comparison we also list the results of VAMPNet embeddings of mFlood.
\begin{table}[h]
    \caption{MSM quality for physical featurizations, scored with the pipeline of \cref{tab:msm-quality} (per domain, leave-one-replica-out at lag $\tau=50$, 10 micro states, TICA onto 7 components). \mfloodph\ is listed for comparison.} \label{tab:app:msm}
    \centering
    \begin{small}
    \begin{tabular}{lrrr}
    \toprule
    feature & \texttt{vamp-2} $\uparrow$ & \texttt{CK RMSE} $\downarrow$ & $\pi$-\texttt{JS} $\downarrow$ \\
    \midrule
    C$_\alpha$ coordinates & 1.392 \std{.002} & 0.040 \std{.000} & 0.254 \std{.001} \\
    Torsion &  1.264 \std{.002} & 0.076 \std{.000} & 0.139 \std{.001} \\
    ($\phi$, $\psi$) + C$_\alpha$ & 1.335 \std{.003} & 0.063 \std{.000} & 0.185 \std{.001} \\
    SASA & 1.439 \std{.002} & 0.034 \std{.000} & 0.258 \std{.001} \\
    \midrule
    \mfloodph & 1.503 \std{.010} & 0.034 \std{.001}  & 0.120 \std{.004}  \\
    \bottomrule
    \end{tabular}
    \end{small}
\end{table}

When using distance based features (see \cref{tab:app:msm}), the covariance matrices become singular on domains with more than 66 residues. On the remaining $\sim$13\% of domains, performance is poor and no real dynamic information is extracted.

\begin{table}[h]
    \caption{MSM quality for TICA projections of distance-based featurizations. These are computable only for the smallest domains, so \emph{\% ok} gives the share of the 532 validation domains that could be scored. Mean over those domains.} \label{tab:app:msm_dist}
    \centering
    \begin{small}
    \begin{tabular}{lrrrr}
    \toprule
    feature & \% ok & \texttt{vamp-2} $\uparrow$ & \texttt{CK RMSE} $\downarrow$ & $\pi$-\texttt{JS} $\downarrow$ \\
    \midrule
    resdist & 12.8 & 1.227 \std{.003} & 0.157 \std{.002} & 0.046 \std{.002} \\
    $d^{-1}$ & 12.8 & 1.225 \std{.002} & 0.159 \std{.001} & 0.057 \std{.001} \\
    $d^{-2}$ & 12.8 & 1.222 \std{.003} & 0.157 \std{.001} & 0.062 \std{.001} \\
    $\log d$ & 12.8 & 1.225 \std{.001} & 0.159 \std{.002} & 0.050 \std{.001} \\
    $e^{-d}$ & 12.8 & 1.216 \std{.001} & 0.158 \std{.001} & 0.079 \std{.001} \\
    $h_4$ & 12.8 & 1.226 \std{.002} & 0.151 \std{.002} & 0.080 \std{.003} \\
    $h_5$ & 12.8 & 1.228 \std{.003} & 0.151 \std{.002} & 0.075 \std{.002} \\
    $h_6$ & 12.8 & 1.222 \std{.003} & 0.156 \std{.002} & 0.065 \std{.001} \\
    $h_8$ & 12.8 & 1.223 \std{.002} & 0.160 \std{.002} & 0.057 \std{.001} \\
    $h_{10}$ & 12.8 & 1.221 \std{.001} & 0.156 \std{.002} & 0.049 \std{.001} \\
    ($e^{-d}$, torsions) & 4.7 & 1.207 \std{.003} & 0.196 \std{.003} & 0.047 \std{.001} \\
    \bottomrule
    \end{tabular}
    \end{small}
\end{table}

In \cref{tab:app:msm_dist}, $d$ refers to the minimal distance between heavy atom pairs of different residues,
excluding neighbors $i+1$ and $i+2$. The contact features are defined as 
\begin{equation}
    h_c(d) = \begin{cases} 
        1, & \text{if} \ d < c \\
        0, & \text{if} \ d \geq c 
    \end{cases}
\end{equation}
for $c \in \{4, 5, 6, 8, 10\}$ in units of \AA.
\FloatBarrier

\subsubsection{mdCATH}
\label[appendix]{subsection:app:mdcathgen}

In \cref{tab:app:mdcathgen} we extend \cref{table:mainclean} and report all ensemble metrics as were reported in
\citet[Table 2 \& Table 7]{Kapusniak.2026}.
In \cref{tab:app:foldedness} we supply additional validity metrics (as in \citet[Table 8]{Kapusniak.2026}) regarding
the ability of the generated ensemble to recover (1) the reference distribution
of native contacts (FNC JSD), (2) $C_\alpha$-level similarity and (3) backbone geometry (GDT-TS JSD, Bond and Angle RMSZ).
We report \marsfm\ results from using the released checkpoint, as $^\dagger$\marsfm\ our own re-training (using the same protocol and hyperparameters) and as $^\ddagger$\marsfm\ the values stated in \citet{Kapusniak.2026}.

\begin{table}[h]
\caption{Complete generative evaluation metrics. 
\texttt{MD oracle} represents the performance of \emph{one} against the distribution of all 5 replicas, and, as baseline, \marsfm\texttt{-random},
lists results of a \marsfm\ model when the guiding MSMs are fitted to 2D Gaussian random features across training trajectories. For each model, we report mean \std{1 std. dev.} over 5 evaluation runs.}
\label{tab:app:mdcathgen}
\centering
\begin{small}
\begin{tabular}{l *{5}{c}}
\toprule
 & \makecell{Pairwise\\RMSD r $\uparrow$} & \makecell{Global\\RMSF r $\uparrow$} & \makecell{Per-target\\RMSF r $\uparrow$} & RG KL $\downarrow$ & RG JSD $\downarrow$ \\
\midrule
\texttt{MD oracle}    & 0.82 \std{.014} & 0.82 \std{.012} & 0.90 \std{.005} & 0.58 \std{.020} & 0.07 \std{.001} \\
\texttt{\marsfm -\texttt{random}}     & 0.57 \std{.004} & 0.68 \std{.003} & 0.89 \std{.001} & 0.51 \std{.002} & 0.10 \std{.000} \\
\midrule
$^\dagger$\marsfm      & 0.66 \std{.003} & 0.72 \std{.002} & 0.89 \std{.003} & 0.55 \std{.002} & 0.10 \std{.000} \\
$^\ddagger$\marsfm      & 0.65 \std{.004} & 0.71 \std{.003} & 0.89 \std{.001} & 0.55 \std{.002} & 0.10 \std{.001} \\
\midrule
\marsfm   & 0.64 \std{.002} & 0.71 \std{.002} & 0.90 \std{.002} & 0.55 \std{.002} & 0.10 \std{.000} \\
\marsfmmflood      & 0.68 \std{.003} & 0.74 \std{.001} & 0.90 \std{.002} & 0.58 \std{.003} & 0.10 \std{.000} \\
\marsfmflood & 0.66 \std{.002} & 0.73 \std{.001} & 0.89 \std{.002} & 0.58 \std{.002} & 0.10 \std{.000} \\
\marsfmalphaca & 0.67 \std{.002} & 0.72 \std{.001} & 0.89 \std{.001} & 0.60 \std{.003} & 0.10 \std{.000} \\
\marsfmalpha & 0.68 \std{.004} & 0.74 \std{.002} & 0.90 \std{.001} & 0.59 \std{.002} & 0.10 \std{.000} \\

\midrule
\midrule

 & SSF KL $\downarrow$ & SSF JSD $\downarrow$ & MSM JSD $\downarrow$ & $\Delta$G MAE $\downarrow$ & $\Delta$G r $\uparrow$ \\
\midrule
\texttt{MD oracle} & 0.61 \std{.048} & 0.07 \std{.003} & 0.19 \std{.006} & 0.90 \std{.046} & 0.90 \std{.002} \\
\texttt{\marsfm -\texttt{random}} & 1.52 \std{.009} & 0.18 \std{.001} & 0.16 \std{.002} & 1.36 \std{.007} & 0.86 \std{.004} \\
\midrule
$^\dagger$\marsfm   & 0.91 \std{.006} & 0.13 \std{.000} & 0.18 \std{.001} & 1.05 \std{.005} & 0.87 \std{.000} \\
$^\ddagger$\marsfm  & 0.93 \std{.010} & 0.13 \std{.001} & 0.19 \std{.004} & 1.05 \std{.003} & 0.87 \std{.003} \\
\midrule
\marsfm    & 1.05 \std{.007} & 0.14 \std{.000} & 0.18 \std{.001} & 0.99 \std{.009} & 0.89 \std{.002} \\
\marsfmmflood   & 0.81 \std{.005} & 0.12 \std{.000} & 0.18 \std{.004} & 0.96 \std{.012} & 0.89 \std{.001} \\
\marsfmflood   & 0.81 \std{.007} & 0.12 \std{.000} & 0.18 \std{.004} & 0.97 \std{.011} & 0.88 \std{.003} \\
\marsfmalphaca        & 0.78 \std{.006} & 0.12 \std{.000} & 0.18 \std{.002} & 0.91 \std{.004} & 0.89 \std{.002} \\
\marsfmalpha        & 0.85 \std{.007} & 0.12 \std{.000} & 0.18 \std{.003} & 0.92 \std{.005} & 0.89 \std{.002} \\
\bottomrule
\end{tabular}
\end{small}
\vspace{-0.2cm}
\end{table}

\begin{table}[h]
\caption{Comparisons on foldedness distributions (FNC JSD), $C_\alpha$ level similarity (GDT-TS JSD) and backbone covalent geometry (Bond / Angle RMSZ).
\label{tab:app:foldedness}}
\vskip0.4ex
\centering
\begin{small}
\begin{tabular}{lrrrr}
\toprule
 & FNC JSD & GDT-TS JSD & Bond RMSZ & Angle RMSZ \\
\midrule
\texttt{MD oracle}         & 0.13 \std{.003} & 0.11 \std{.004} & 2.23 \std{.012} & 2.19 \std{.009} \\
\texttt{MarS-FM-random}                 & 0.17 \std{.001} & 0.14 \std{.000} & 6.99 \std{.048} & 3.14 \std{.003} \\
\midrule
$^\dagger$\marsfm\                       & 0.14 \std{.001} & 0.11 \std{.000} & 6.95 \std{.071} & 3.12 \std{.003} \\
$^\ddagger$\marsfm\  & 0.15 \std{.001} & 0.12 \std{.001} & 12.62 \std{.144} & 3.06 \std{.001} \\
\midrule
\marsfm\                                & 0.15 \std{.000} & 0.12 \std{.000} & 7.04 \std{.055} & 3.16 \std{.005} \\
\marsfmmflood                           & 0.14 \std{.001} & 0.11 \std{.000} & 5.20 \std{.018} & 2.94 \std{.003} \\
\marsfmflood                            & 0.14 \std{.000} & 0.11 \std{.000} & 5.07 \std{.021} & 2.95 \std{.005} \\
\marsfmalphaca                            & 0.14 \std{.000} & 0.11 \std{.000} & 5.26 \std{.057} & 2.90 \std{.004} \\
\marsfmalpha                            & 0.14 \std{.001} & 0.11 \std{.000} & 6.21 \std{.050} & 3.12 \std{.003} \\
\hline
\end{tabular}
\end{small}
\end{table}

To establish statistical significance for domain-wise metrics\footnote{For metrics comparing the whole generated ensemble (Pairwise RMSD r, Global RMSF r, $\Delta$G r) no individual domain values exist, such that these cannot be compared in this manner.}, 
we compare the original \marsfm\ with \marsfmmflood\ using two-sided paired t-tests, followed by
Benjamini–Hochberg correction across all 11 per-domain metrics (q=0.05) in \cref{tab:app:gensignificance}. The \marsfmmflood\ variant performs significantly
better on five metrics, i.e., SSF KL, SSF JSD, bond RMSZ, angle RMSZ, and per-target RMSF correlation,
whereas the original \marsfm\ performs significantly better on RG KL.
No statistically significant difference is detected for the remaining five metrics. 
The \emph{Direction of mean difference} indicates which method yielded the better result in terms of absolute numbers.

\begin{table}[h]
\caption{%
    Significance assessment of generated ensemble statistics between \marsfm\ and \marsfmmflood.%
    \label{tab:app:gensignificance}
}
\begin{tabular}{l rrr}
    \toprule
    Metric & Direction of mean difference & FDR-adjusted p-value & Significant? \\
    \midrule
    RG KL & \marsfm &0.0002 & Yes \\
    RG JSD & \marsfm &0.8607 & No \\
    SSF KL & \marsfmmflood &1e-10 & Yes \\
    SSF JSD & \marsfmmflood &1e-10 & Yes \\
    MSM JSD & \marsfmmflood &0.8607 & No \\
    dG MAE & \marsfmmflood &0.1577 & No \\
    FNC JSD & \marsfmmflood &0.2959 & No \\
    GDT-TS JSD & \marsfmmflood &0.9428 & No \\
    Bond RMSZ & \marsfmmflood &1e-10 & Yes \\
    Angle RMSZ & \marsfmmflood &1e-10 & Yes \\
    Per-target RMSF r & \marsfmmflood &0.0435 & Yes \\
    \bottomrule
\end{tabular}
    
\end{table}

\FloatBarrier
\subsection{Consistency across evaluations}
\label[appendix]{subsection:app:ranks}
\Cref{tab:app:ranks} summarizes how the PH-based representations rank against each other across the evaluations of the main text, using all generative metrics of \cref{subsection:app:mdcathgen}.
For each metric, we rank them at the precision reported in the respective table, with ties sharing the better rank, and average the ranks per evaluation.

\begin{table}[h]
\caption{Mean rank of the PH-based representations per evaluation (1 = best): CATH classification (2 metrics, \cref{table:clf}), observable regression (8 metrics, \cref{table:regression}), MSM quality (3 metrics, \cref{tab:msm-quality}) and generation (14 metrics, \cref{tab:app:mdcathgen,tab:app:foldedness}; \ripsphca\ not evaluated). \emph{Overall} and \emph{Worst} give the mean and the worst rank over all metrics; \emph{Overall w/o generation} averages over the 13 classification, regression and MSM metrics, on which all five representations are evaluated.\label{tab:app:ranks}}
\centering
\begin{small}
\begin{tabular}{lccccccc}
\toprule
 & Classification & Regression & \makecell{MSM\\quality} & Generation & Overall & \makecell{Overall w/o\\generation} & Worst \\
\midrule
\mfloodph & \textbf{1.50} & 2.25 & \textbf{1.67} & \textbf{1.36} & \textbf{1.67} & \textbf{2.00} & \textbf{3} \\
\floodph & 2.50 & \textbf{2.00} & 3.33 & 2.14 & 2.26 & 2.38 & 4 \\
\alphaph & 5.00 & 2.63 & 2.33 & 1.86 & 2.37 & 2.92 & 5 \\
\alphaphca & 3.00 & 3.25 & 3.33 & 1.86 & 2.52 & 3.23 & 5 \\
\ripsphca & 2.50 & 4.88 & 4.00 & -- & 4.31 & 4.31 & 5 \\
\bottomrule
\end{tabular}
\end{small}
\end{table}

\subsection{Fast Folders}
\label[appendix]{subsection:app:fastfolders}
\Cref{tab:rmsd_fnc_gdt} compares native-state conformations generated by \marsfmmflood\ with finetuning to those of the MD reference, measuring RMSD, FNC and GDT-TS with respect to the experimental structure, with the native state defined as in \cref{subsection:app:datafastfolders}. For proteins with a tight native basin in the reference (reference RMSD below 2~\AA, i.e., Chignolin, Trp-Cage, WW-Domain, Villin, NTL9 and Protein G), generated native conformations deviate from the crystal structure by 0.6 to 1.6~\AA\ more than the reference. For proteins whose native basin is already broad in the reference, this gap narrows and nearly vanishes for BBA, BBL and Homeodomain, indicating that larger deviations there reflect the breadth of the native ensemble rather than a failure of the model.

\begin{table}[h]
\centering
\caption{Mean RMSD, FNC, and GDT-TS for generated and MD reference trajectories (fast folders) all between structures
clustered into the native state with respect to the experimental crystal structure.}
\vskip0.4ex
\label{tab:rmsd_fnc_gdt}
\setlength{\tabcolsep}{5pt}
\small
\begin{tabular}{l ccc ccc}
\toprule
& \multicolumn{3}{c}{Generated} & \multicolumn{3}{c}{Reference} \\
\cmidrule(lr){2-4} \cmidrule(lr){5-7}
Protein
& \makecell{Mean\\RMSD (\AA)} & FNC & GDT-TS
& \makecell{Mean\\RMSD (\AA)} & FNC & GDT-TS \\
\midrule
Chignolin           & 1.58 \std{0.53} & 0.70 \std{0.23} & 0.80 \std{0.10} & 0.95 \std{0.33} & 0.91 \std{0.10} & 0.92 \std{0.07} \\
Trp-Cage            & 2.24 \std{0.72} & 0.85 \std{0.10} & 0.67 \std{0.10} & 1.57 \std{0.37} & 0.91 \std{0.05} & 0.76 \std{0.07} \\
BBA                 & 4.08 \std{1.27} & 0.76 \std{0.11} & 0.52 \std{0.11} & 3.60 \std{1.27} & 0.74 \std{0.09} & 0.58 \std{0.11} \\
WW-Domain           & 3.55 \std{1.16} & 0.74 \std{0.09} & 0.55 \std{0.10} & 1.97 \std{0.55} & 0.90 \std{0.06} & 0.73 \std{0.08} \\
Villin              & 2.74 \std{0.82} & 0.86 \std{0.06} & 0.66 \std{0.11} & 1.76 \std{0.60} & 0.94 \std{0.04} & 0.82 \std{0.08} \\
NTL9                & 1.88 \std{0.56} & 0.81 \std{0.09} & 0.75 \std{0.09} & 0.85 \std{0.34} & 0.96 \std{0.05} & 0.95 \std{0.05} \\
BBL                 & 4.92 \std{1.40} & 0.68 \std{0.09} & 0.45 \std{0.10} & 5.01 \std{1.53} & 0.67 \std{0.09} & 0.44 \std{0.12} \\
Protein B           & 4.05 \std{1.23} & 0.61 \std{0.09} & 0.48 \std{0.10} & 3.30 \std{1.36} & 0.65 \std{0.08} & 0.56 \std{0.12} \\
Homeodomain         & 3.98 \std{1.38} & 0.79 \std{0.08} & 0.57 \std{0.12} & 3.96 \std{1.41} & 0.82 \std{0.08} & 0.61 \std{0.14} \\
Protein G           & 2.43 \std{0.76} & 0.82 \std{0.09} & 0.66 \std{0.10} & 1.63 \std{0.62} & 0.91 \std{0.06} & 0.82 \std{0.09} \\
$\alpha$3D          & 3.87 \std{0.92} & 0.70 \std{0.06} & 0.48 \std{0.08} & 2.91 \std{0.49} & 0.69 \std{0.04} & 0.57 \std{0.05} \\
$\lambda$-repressor & 6.45 \std{1.60} & 0.54 \std{0.07} & 0.34 \std{0.09} & 5.48 \std{1.44} & 0.56 \std{0.08} & 0.40 \std{0.10} \\
\bottomrule
\end{tabular}
\end{table}

\begin{figure}[t!]
\centering{
    \includegraphics[width=0.90\textwidth]{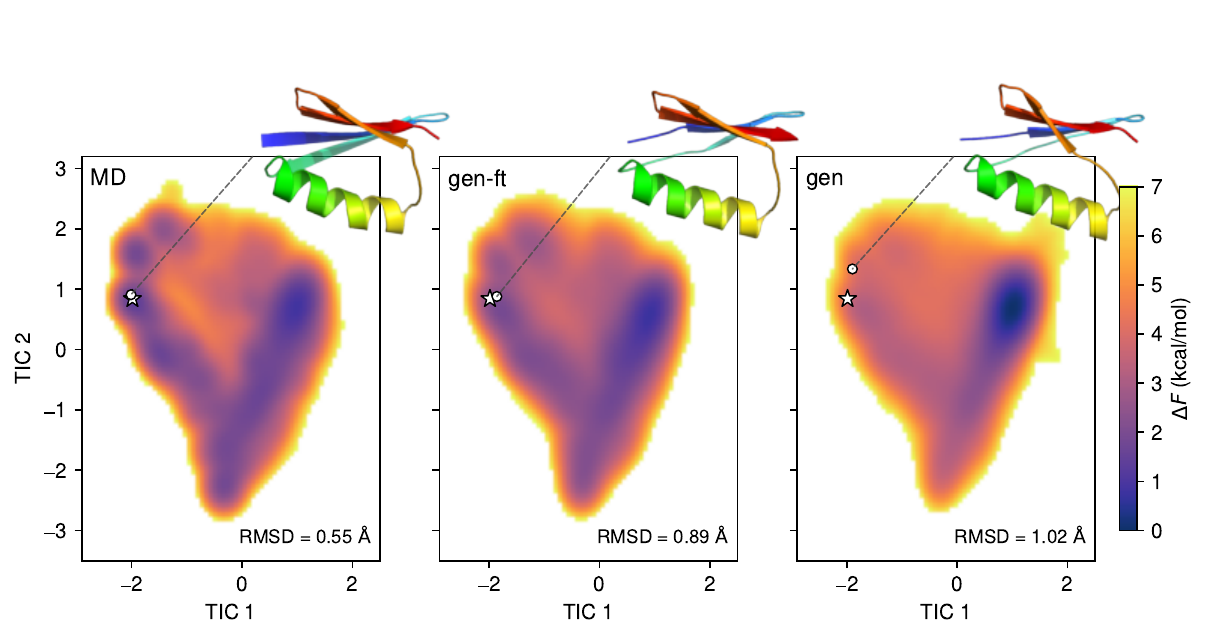}}
    \caption{Free energy landscapes of Protein G in TICA coordinates ($\tau=10$ ns),
    comparing the MD reference trajectories (\textbf{Left}, \texttt{MD})
    to samples generated by \marsfmmflood\ with finetuning (\textbf{Middle},
    \texttt{gen-ft}) and without finetuning (\textbf{Right}, \texttt{gen}).
    Cartoon insets show the frames closest in RMSD 
    to the experimental crystal structure (1MI0, marked by a star);
    the corresponding TIC coordinates are marked by a circle, and the per-panel
    $C_\alpha$ RMSD to the crystal structure is annotated in the lower right.
    We clamp $\Delta F$, the difference to the lowest basin, at 7 kcal/mol.}
    \label{fig:app:ff_proteing_landscape}
\end{figure}

\Cref{fig:app:ff_proteing_landscape} illustrates the effect of finetuning for Protein G. Without finetuning, most generated probability mass falls into a single basin distant from the crystal structure, and the native region is only sparsely populated. After finetuning, the generated landscape recovers the overall shape of the MD reference, including the basins near the native state, and the generated frame closest to the crystal structure improves from 1.02~\AA\ to 0.89~\AA, compared to 0.55~\AA\ in the reference.

\cref{tab:app:ff_rmsz} shows that covalent geometry is of comparable quality across all proteins. Bond RMSZ of generated conformations ranges from 2.6 (Homeodomain) to 3.7 ($\lambda$-repressor) and angle RMSZ from 2.1 to 2.7, compared to 1.9 to 2.0 for the MD reference. 
The mean is raised by a small fraction of strongly distorted frames: the median bond RMSZ lies between 2.3 and 2.8, and 95.9\% ($\lambda$-repressor) to 99.7\% (Chignolin) of generated frames stay below a bond RMSZ of 10. This fraction of distorted frames is among the lowest for Chignolin and Trp-Cage, the proteins for which \citet{Costa.2025} report chemically invalid states.

\begin{table}[h]
\centering
\caption{Covalent geometry per fast folding protein for conformations generated by \marsfmmflood\ with finetuning and for the MD reference, computed over all frames. Besides mean RMSZ, we report the median bond RMSZ and the percentage of generated frames with bond RMSZ above 10, i.e., far beyond the maximum of 3.5 observed in the MD reference. We omit both for angle RMSZ, which shows no comparable tail (99th percentile at most 6.3). Averages over proteins are reported in \cref{table:ff_validation}.}
\vskip0.4ex
\label{tab:app:ff_rmsz}
\small
\begin{tabular}{l cccc cc}
\toprule
& \multicolumn{4}{c}{Generated} & \multicolumn{2}{c}{Reference} \\
\cmidrule(lr){2-5} \cmidrule(lr){6-7}
& \multicolumn{3}{c}{Bond RMSZ} & Angle RMSZ & Bond RMSZ & Angle RMSZ \\
\cmidrule(lr){2-4}
Protein & mean & median & \% $>10$ & mean & mean & mean \\
\midrule
Chignolin           & 2.92 & 2.65 & 0.33 & 2.51 & 1.95 & 1.87 \\
Trp-Cage            & 3.23 & 2.81 & 0.93 & 2.70 & 2.00 & 1.99 \\
BBA                 & 2.95 & 2.62 & 0.78 & 2.31 & 1.96 & 1.84 \\
WW-Domain           & 3.21 & 2.77 & 1.24 & 2.52 & 1.97 & 1.89 \\
Villin              & 3.08 & 2.59 & 1.38 & 2.34 & 1.93 & 1.86 \\
NTL9                & 3.08 & 2.60 & 1.42 & 2.27 & 1.96 & 1.83 \\
BBL                 & 3.48 & 2.67 & 2.58 & 2.40 & 1.99 & 1.86 \\
Protein B           & 2.95 & 2.43 & 1.57 & 2.21 & 1.93 & 1.84 \\
Homeodomain         & 2.62 & 2.29 & 0.96 & 2.12 & 1.92 & 1.84 \\
Protein G           & 3.35 & 2.58 & 2.59 & 2.29 & 1.96 & 1.85 \\
$\alpha$3D          & 3.35 & 2.51 & 2.94 & 2.27 & 1.98 & 1.86 \\
$\lambda$-repressor & 3.69 & 2.57 & 4.07 & 2.31 & 1.99 & 1.86 \\
\bottomrule
\end{tabular}
\end{table}

Since no model is trained on fast folding proteins, and the finetuned variants are adapted on the 348~K subset of \mdcath\ only (\cref{subsection:app:implfastfolders}), both the close agreement for broad native basins and the recovered Protein G landscape are obtained on proteins not seen during training.

\FloatBarrier

\section{Properties of the masked Flood complex}
\label[appendix]{section:app:theory}
\subsection{Bottleneck stability}
\begin{theorem}
    Let $L\subset \mathbb R^d$ be finite, and for each $\sigma \in \Del(L)$ let $X_\sigma, X'_\sigma \subset \mathbb R^d$ be finite such that $X_\sigma\subset X_\tau$ and $X'_\sigma \subset X'_\tau$ for all $\tau\subset \sigma$.
    The masked Flood complex then satisfies the following stability result for all $i \ge 0$:
    \begin{equation}
        d_B\big(
        \dgm_i(\mFlood(\{X_\sigma\}_{\sigma},L)), \dgm_i(\mFlood(\{X'_\sigma\}_{\sigma},L))
        \big)
        \le
        \max_{\sigma \in \Del(L)} d_H(X_\sigma, X'_\sigma)
        \enspace.
    \end{equation}
\end{theorem}
\begin{proof}
    Let $\sigma \in \Del(L)$ and let $f(\sigma), f'(\sigma)$ denote its respective filtration values.
    The triangle inequality for directed Hausdorff distances implies
    \begin{equation}
        f(\sigma) 
        = \max_{p\in \hull(\sigma)} d(p, X_\sigma)
        \le \max_{p\in \hull(\sigma)} d(p, X'_\sigma) + \max_{x'\in X'_\sigma} d(x', X_\sigma)
        = f'(\sigma) + \max_{x'\in X'_\sigma} d(x', X_\sigma)
        \enspace,
    \end{equation}
    and analogously, $f'(\sigma) \le f(\sigma) + \max_{x\in X_\sigma} d(x, X'_\sigma)$. In sum, we have
    \begin{equation}
        |f(\sigma) - f'(\sigma)| 
        \le \max\big\{\max_{x\in X_\sigma} d(x, X'_\sigma), \max_{x'\in X'_\sigma} d(x', X_\sigma)\big\}
        = d_H(X_\sigma, X'_\sigma)
        \enspace.
    \end{equation}
    Taking the maximum over all $\sigma \in \Del(L)$ gives a uniform bound, so that we can apply the stability theorem from \cite{Cohen-Steiner2007a}, which then finishes the proof.
\end{proof}

\subsection{Computation from grid}
As in \cite{Graf.2026} we approximate the filtration values of $\mFlood$ using a grid of evenly spaced barycentric coordinates. Given a predefined number $m\in \mathbb N$ controlling the number of grid points along each edge, we denote the resulting grid points on a simplex $\sigma$ by $P_\sigma = \{\sum_{i=0}^k \lambda_i v_i \colon \lambda_i \ge 0, \sum_{i=0}^k \lambda_i = 1, m \lambda_i \in \mathbb N_{\ge0} \} \subset \hull{\sigma}$. 
We then define $\mFlood_r(\{X_\sigma\}_\sigma, L , \{P_\sigma\}_\sigma )$ as 
\begin{equation}
    \mFlood_r(\{X_\sigma\}_\sigma, L , \{P_\sigma\}_\sigma )
    = \bigg\{\sigma \in \Del(L) \colon P_\sigma \subset \bigcup_{x\in X_\sigma} B_r(x) \bigg\}
    \enspace.
\end{equation}
Equivalently, $\mFlood$ is the sublevel set of the function 
\begin{equation}
    f: \Del(L) \to \mathbb R \qquad \sigma \mapsto \max_{p\in P_\sigma} \min_{x\in X_\sigma} d(p,x)
\end{equation}
which, as opposed to the non-convex optimization problem of finding the filtration values of the exact $\mFlood(\{X_\sigma\}_\sigma, L)$ can be computed directly from the pairwise distances between $P_\sigma$ and $X_\sigma$.
For all experiments, we use a resolution of $m=20$ grid points per edge.

As we show below, $f$ is a well-defined filtration function satisfying $f(\tau) \le f(\sigma)$ whenever $\tau\subset \sigma$, and thus the family $\mFlood_r(\{X_\sigma\}_\sigma, L , \{P_\sigma\}_\sigma )$ for $r\ge 0$ forms a filtration of $\Del(L)$.
Indeed, if $\tau\subset \sigma$, then $P_\tau\subset P_\sigma$, and thus 
\begin{align}
        f(\tau) 
         = \max_{p\in P_\tau} \min_{x\in X_{\tau}} d(p,x) 
         \le \max_{p\in P_\sigma} \min_{x\in X_{\tau}} d(p,x) 
         \le \max_{p \in P_\sigma} \min_{x\in X_{\sigma}} d(p,x) 
         = f(\sigma)
        \enspace.
\end{align}

Moreover, we have the following approximation guarantee.

\begin{proposition}
    Let $L\subset \mathbb R^d$ be a finite set, and for each $\sigma \in \operatorname{Del}(L)$ let $X_\sigma \subset \mathbb R^d$ be finite such that $X_\sigma \subset X_\tau$ for all $\tau \subset \sigma$, and given $m\in \mathbb N_{>0}$, let $P_\sigma = \{\sum_{i=0}^k \lambda_i v_i \colon \lambda_i \ge 0, \sum_{i=0}^k \lambda_i = 1, m \lambda_i \in \mathbb N_{\ge0} \} \subset \hull{\sigma}$.
    Then, denoting $g(\{v_0,\dots,v_k\}) := \sqrt{\sum_{0\le a < b \le k} \|v_a -v_b \|^2}$, we have for every $i\ge 0$
    \begin{equation}
        d_B(
        \dgm_i(\mFlood(\{X_\sigma\}_\sigma, L)),
        \dgm_i(\mFlood(\{X_\sigma\}_\sigma, L , \{P_\sigma\}_\sigma))
        )
        \le 
        \frac 1 m \max_{\sigma \in \Del(L)} g(\sigma)
        \enspace.
    \end{equation}
\end{proposition}
\begin{proof}
    We note that \cite[Theorem~5]{Graf.2026} already shows $d_B\left(\dgm_i(\mathrm{Flood}(X,L)), \dgm_i(\mathrm{Flood}(X,L,P))\right) \le \max_{\sigma \in \mathrm{Del}(L)} d_H(P_\sigma, \hull{\sigma})$ for the original Flood complex, and \cite[Lemma~6]{Graf.2026} shows that 
    $d_H(P_{\tau}, \hull{\tau}) \le \frac 1 m \sqrt{\sum_{i<j} \Vert v_i - v_j \Vert^2} = g(\tau)/m$ for any simplex $\tau$.
    We therefore only need to show that \cite[Theorem~5]{Graf.2026} extends to the masked Flood complex.

    Indeed, it is clear from the definitions that $\tau \in  \mFlood_r(\{X_\sigma\}_\sigma, L)$ implies $\tau \in  \mFlood_r(\{X_\sigma\}_\sigma, L, \{P_\sigma\}_\sigma)$ since $P_\tau \subset \hull \tau$.
    Conversely, let $\tau \in \mFlood_r(\{X_\sigma\}_\sigma, L, \{P_\sigma\}_\sigma)$.
    Thus, for each $p \in P_\tau$ there is $x_p \in X_\tau$ with $d(p, {x_p}) \le r$. Moreover, with $\varepsilon := \max_{\sigma \in \Del(L)} d_H(P_\sigma, \hull \sigma)$ there is for any $q\in \hull \tau$ some $p' \in P_\tau$ with $d(q, p') \le \varepsilon$.
    Hence, 
    \begin{equation*}
        d(q, X_\tau) \le d(q,x_{p'}) \le d(q, p') + d(p', x_{p'}) \le r + \varepsilon
        \enspace.
    \end{equation*}
    which implies $\mFlood_r(\{X_\sigma\}_\sigma, L, \{P_\sigma\}_\sigma) \subset \mFlood_{r+\varepsilon}(\{X_\sigma\}_\sigma, L)$.
    The classic stability theorem from \cite{Cohen-Steiner2007a} then implies 
    \begin{align*}
        d_B &\left(\dgm_i(\mFlood(\{X_\sigma\}_\sigma,L)), \dgm_i(\mFlood(\{X_\sigma\}_\sigma,L,\{P_\sigma\}_\sigma))\right) 
        \\
        & \le \max_{\sigma \in \mathrm{Del}(L)} d_H(P_\sigma, \hull{\sigma})
        \le  \frac 1 m  \max_{\sigma\in \Del(L)} g(\sigma)
        \enspace.
    \end{align*}
\end{proof}

\section{Runtime and memory}
\label{sec:runtime}

We performed runtime and peak-memory profiling for the PH computation ($H_0$,$H_1$,$H_2$), using 500 randomly selected \mdcath\ domains for runtime and the ten largest domains for memory. 
\floodph\ is fastest at 14.4 ms per frame; \mfloodph\ requires 28.4 ms, comparable to edge-collapsed \ripsphca\ at 25.0 ms (but Rips scales substantially worse with protein size), and $3.3\times$ faster than \alphaph\ at 94.1 ms. All-atom Rips was infeasible at this scale. 
The overall cost on \mdcath\ arises from its large number of conformations, i.e., $\sim$12.5M conformations ($\sim$5k domains with 5 replicas with $\sim$500 time steps), rather than individual protein size (maximal 500 C$_\alpha$, $\sim$8000 atoms). 
Due to the low memory requirements, PH computation is parallelizable across domains (trivially by starting one process per domain), which reduces total wall time by an order of magnitude. Moreover, batching over simplices (as is default in the Flooder implementation) reduces \mfloodph\ peak GPU memory from 1,185 MB to 384 MB without affecting runtime, while its approximately 1.5 GB host-memory footprint is dominated by fixed CUDA/Triton process overhead. 
The total wall time column lists the estimate without any parallelization across systems (which is trivially doable).

\begin{table}[t]
\centering
\caption{Comparison of runtime and memory usage for PH computation in degrees 0 -- 2.}
\label{tab:performance}
\begin{small}
\begin{tabular}{llrrrr}
\toprule
Method & Batch & \makecell{Wall / diagram\\(ms)} & 
\makecell{Est. total\\wall time (h)} & 
\makecell{Peak GPU\\memory (MB)} & 
\makecell{Peak host\\memory (MB)} \\
\midrule
\ripsphca & ---  & $25.0 \pm 56.0$  & 79  & ---   & 333  \\
\alphaph  & ---  & $94.1 \pm 63.5$  & 301 & ---   & 409  \\
\floodph  &1024 & $14.4 \pm \phantom{0}7.1$   & 46  & 93    & 1548 \\
\floodph  &4096 & $14.9 \pm 10.5$  & 48  & 164   & 1550 \\
\mfloodph &1024 & $28.4 \pm 21.1$  & 91  & 384   & 1528 \\
\mfloodph &4096 & $28.4 \pm 21.0$  & 91  & 1185  & 1516 \\
\bottomrule
\end{tabular}
\end{small}
\end{table}

In terms of end-to-end training cost to obtain vectorizations, the $k$-means++ clustering on the points in the persistence diagram can be done in less than 10 min \citep{douze2024faiss}
(using FAISS on the GPU), and VAMPNet training at corpus-scale in less than 20 min on our system (NVIDIA H100 GPU, see \cref{app:compute}).

For generative modeling, once VAMPNet features are pre-computed for all training MD trajectories, no additional computational
cost is added to \marsfm.

\section{Ablations}
\label[appendix]{section:app:ablations}

\subsection{Homology degree}

We report which homology dimensions provide relevant information for the CATH-classification experiment. 
Indeed, higher-dimensional features do provide additional 
information, highlighting the relevance for methods (such as our 
masked Flood complex) that enable efficiently computing the latter.

\begin{table}[t]
    \centering    
    \caption{Ablation of homology degree contribution to CATH classification performance.}
    \begin{small}
    \begin{tabular}{lcc}
        \toprule
        & \textbf{Linear} & \textbf{MLP} \\
        \midrule
        \mfloodph\ ($H_0$)             & $0.692 \pm 0.002$ & $0.679 \pm 0.009$ \\
        \mfloodph\ ($H_1$)             & $0.767 \pm 0.015$ & $0.766 \pm 0.023$ \\
        \mfloodph\ ($H_2$)             & $0.752 \pm 0.002$ & $0.796 \pm 0.029$ \\
        \mfloodph\ ($H_0$, $H_1$)      & $0.767 \pm 0.003$ & $0.785 \pm 0.011$ \\
        \mfloodph\ ($H_0$, $H_2$)      & $0.761 \pm 0.012$ & $0.802 \pm 0.004$ \\
        \mfloodph\ ($H_1$, $H_2$)      & $0.789 \pm 0.017$ & $0.806 \pm 0.033$ \\
        \mfloodph\ ($H_0$, $H_1$, $H_2$) & $0.791 \pm 0.007$ & $0.812 \pm 0.021$ \\
        \bottomrule
    \end{tabular}
    \end{small}
    \label{tab:mflood_results}
\end{table}

\subsection{Masking strategy}

In the table below, we compare the current masking variant, \mfloodph\ (nbs0), which suppresses within-residue interactions while retaining longer-range geometry, with \mfloodph\ (nbs1), which also masks interactions between neighboring residues. 
This more aggressive mask leaves classification and regression largely unchanged: balanced accuracy changes from 0.812 $\pm$ 0.021 to 0.819 $\pm$ 0.014, with mixed $R^2$ changes of at most 0.016, but degrades MSM consistency. Although VAMP-2 increases from 1.503 to 1.515, CK RMSE worsens from 0.034 to 0.052 and $\pi$-JS from 0.120 to 0.250, eliminating the gains over the unmasked Flood complex variant. Thus, additional masking is not uniformly beneficial, and the original minimal rule provides the best balance across tasks. This ablation also illustrates why VAMP-2 should not be interpreted in isolation.

\begin{table}[ht]
\centering
\caption{Ablation on neighborhood size (number of residues) in the masking of \mfloodph. Regression probes here use a 5-layer MLP head trained for at most 10 epochs; shared rows therefore differ from \cref{table:regression} by at most 0.005.}
\label{tab:results}
    \begin{subtable}{0.48\textwidth}
        \centering
        \caption{CATH classification (balanced accuracy)}
        \label{tab:cath}
        \begin{small}
        \begin{tabular}{llr}
        \toprule
        Feature & Head & Balanced Acc. \\
        \midrule
        Flood             & mlp & 0.800 \\
        mFlood (nbs0) & mlp & 0.812 \\
        mFlood (nbs1)     & mlp & 0.819 \\
        \bottomrule
        \end{tabular}
        \end{small}
        \end{subtable}
        \hfill
        \begin{subtable}{0.48\textwidth}
        \centering
        \caption{MSM quality assessment}
        \label{tab:msm}        
        \begin{small}
        \begin{tabular}{lrrr}
        \toprule
        Feature & VAMP2 & CK-RMSE & $\pi$-JS \\
        \midrule
        Flood             & 1.482 & 0.051 & 0.219 \\
        mFlood (nbs0) & 1.503 & 0.034 & 0.120 \\
        mFlood (nbs1)     & 1.515 & 0.052 & 0.250 \\
        \bottomrule
        \end{tabular}
        \end{small}
    \end{subtable}
    
    \vspace{1em}
    
    \begin{subtable}{\textwidth}
    \centering
    \caption{Observable regression ($R^2$)}
    \label{tab:regression}
        \begin{small}
        \begin{tabular}{llrrrr}
        \toprule
        Feature & Head & RG & SSF & FNC & RMSD \\
        \midrule
        Flood             & mlp & 0.864 & 0.848 & 0.864 & 0.781 \\
        mFlood (nbs0) & mlp & 0.864 & 0.839 & 0.860 & 0.781 \\
        mFlood (nbs1)     & mlp & 0.848 & 0.851 & 0.858 & 0.770 \\
        \bottomrule
        \end{tabular}
        \end{small}
    \end{subtable}
\end{table}

\subsection{Sensitivity to hyperparameters}
For vectorizing persistence diagrams we need to select two hyperparameters: The number of centers, controlling the number of features the vectorization produces and the stretch parameter that controls how much low-lifetime points are downweighted. We evaluate their effect on three experiments: CATH classification, regression on observables and MSM construction. 
Throughout all experiments, $k$-means++ is used to determine cluster centers (re-initialized 5 times; objective 
function = total squared errors, as implemented in FAISS). 
For the third hyperparameter, the bandwidth, we use half the distance from each center to its closest other center, as in ATOL \citep{Royer.2021}.

When varying the number of centers between 32--256, all results are fairly stable; only once the discretization becomes too coarse (e.g., 16 per homology dimension), results on all tasks deteriorate. Increasing the stretch quantile by an order of magnitude (1\% to 10\%) lowers CATH accuracy from 0.812 to 0.778, while regression and MSM quality are unaffected. 

\begin{table}[ht]
\centering
\caption{
Sensitivity analysis on persistent homology vectorization parameters for CATH classification.
$^*$ denotes the configuration reported in the main part of the manuscript.}
\label{tab:cath_sensitivity}

\begin{subtable}{0.48\textwidth}
\centering
\caption{Number of structure elements}
\begin{tabular}{llr}
\toprule
Feature & Head & Balanced Acc. \\
\midrule
mFlood (16)      & mlp & $0.794 \pm 0.031$ \\
mFlood (32)      & mlp & $0.827 \pm 0.004$ \\
mFlood (64)      & mlp & $0.813 \pm 0.034$ \\
mFlood (128)$^*$ & mlp & $0.812 \pm 0.021$ \\
mFlood (256)     & mlp & $0.807 \pm 0.018$ \\
\bottomrule
\end{tabular}
\end{subtable}
\hfill
\begin{subtable}{0.48\textwidth}
\centering
\caption{Stretch quantile}
\begin{tabular}{llr}
\toprule
Feature & Head & Balanced Acc. \\
\midrule
mFlood (128, 1\%)$^*$ & mlp & $0.812 \pm 0.021$ \\
mFlood (128, 5\%)     & mlp & $0.809 \pm 0.021$ \\
mFlood (128, 10\%)    & mlp & $0.778 \pm 0.015$ \\
\bottomrule
\end{tabular}
\end{subtable}

\end{table}

\begin{table}[ht]
\centering
\caption{
Sensitivity analysis on persistent homology vectorization parameters for observable regression ($R^2$).
$^*$ denotes the configuration reported in the main part of the manuscript. Regression probes here use a 5-layer MLP head trained for at most 10 epochs; the $^*$ rows therefore differ from \cref{table:regression} by at most 0.003.
}
\label{tab:regression_sensitivity}

\begin{subtable}{\textwidth}
\centering
\caption{Number of structure elements}
\begin{small}
\begin{tabular}{llrrrr}
\toprule
Feature & Head & RG & SSF & FNC & RMSD \\
\midrule
mFlood (16)      & mlp & 0.841 & 0.806 & 0.840 & 0.764 \\
mFlood (32)      & mlp & 0.861 & 0.832 & 0.857 & 0.778 \\
mFlood (64)      & mlp & 0.856 & 0.820 & 0.849 & 0.773 \\
mFlood (128)$^*$ & mlp & 0.864 & 0.839 & 0.860 & 0.781 \\
mFlood (256)     & mlp & 0.866 & 0.842 & 0.862 & 0.783 \\
\bottomrule
\end{tabular}
\end{small}
\end{subtable}

\vspace{1em}

\begin{subtable}{\textwidth}
\centering
\caption{Stretch quantile}
\begin{small}
\begin{tabular}{llrrrr}
\toprule
Feature & Head & RG & SSF & FNC & RMSD \\
\midrule
mFlood (128, 1\%)$^*$ & mlp & 0.864 & 0.839 & 0.860 & 0.781 \\
mFlood (128, 5\%)     & mlp & 0.864 & 0.838 & 0.859 & 0.781 \\
mFlood (128, 10\%)    & mlp & 0.864 & 0.837 & 0.858 & 0.780 \\
\bottomrule
\end{tabular}
\end{small}
\end{subtable}

\end{table}

\begin{table}[ht]
\centering
\caption{
Sensitivity analysis for MSM quality assessment.
$^*$ denotes the configuration reported in the manuscript.
}
\label{tab:msm_sensitivity}

\begin{subtable}{0.48\textwidth}
\centering
\caption{Number of structure elements}
\begin{small}
\setlength{\tabcolsep}{3.5pt}
\begin{tabular}{lrrr}
\toprule
Feature & VAMP2 & CK-RMSE & $\pi$-JS \\
\midrule
mFlood (16)      & 1.458 & 0.033 & 0.102 \\
mFlood (32)      & 1.488 & 0.032 & 0.114 \\
mFlood (64)      & 1.498 & 0.032 & 0.118 \\
mFlood (128)$^*$ & 1.503 & 0.034 & 0.120 \\
mFlood (256)     & 1.500 & 0.034 & 0.121 \\
\bottomrule
\end{tabular}
\end{small}
\end{subtable}
\hfill
\begin{subtable}{0.48\textwidth}
\centering
\caption{Stretch quantile}
\begin{small}
\setlength{\tabcolsep}{3.5pt}
\begin{tabular}{lrrr}
\toprule
Feature & VAMP2 & CK-RMSE & $\pi$-JS \\
\midrule
mFlood (128, 1\%)$^*$ & 1.503 & 0.034 & 0.120 \\
mFlood (128, 5\%)     & 1.511 & 0.033 & 0.120 \\
mFlood (128, 10\%)    & 1.501 & 0.033 & 0.121 \\
\bottomrule
\end{tabular}
\end{small}
\end{subtable}

\end{table}

\section{Societal impact}
\label[appendix]{section:app:societalimpact}

By improving MSM-based sampling within the generative \marsfm\ framework, our approach may enable more efficient exploration of protein conformational landscapes, with potential applications in areas such as drug discovery and molecular design. Besides these benefits, considerations regarding societal impact analogous to those discussed in \cite{Kapusniak.2026} apply. For example, increased accessibility and realism of generative protein modeling may lower practical barriers to protein engineering, or encourage unwarranted confidence when applied outside validated regimes. While we regard these concerns legitimate, we currently assess the risk of misuse as relatively low and, therefore, do not identify a need for specific technical safeguards beyond standard responsible research practices.

\vspace{-1.0ex}

\clearpage
\section{Additional figures}
\label[appendix]{section:app:additionalfigures}

\begin{figure}[h!]
\vspace{-0.3cm}
\centering{
    \includegraphics[width=0.89\textwidth]{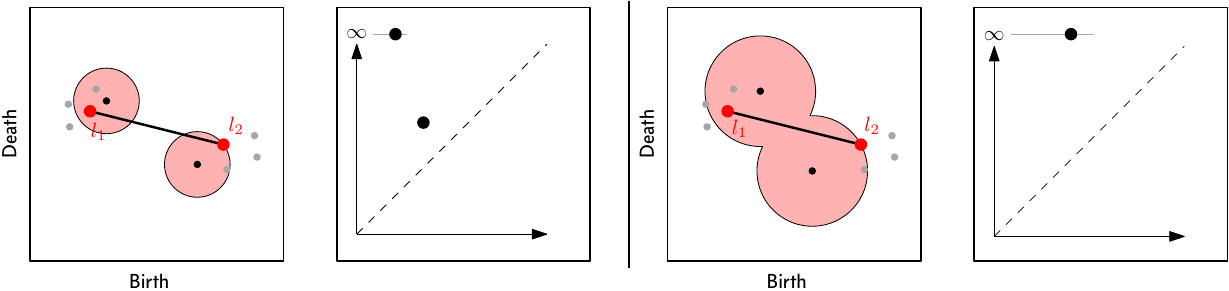}}
    \caption{Filtration values and $H_0$-diagrams of the masked Flood complex. \textcolor{red}{Red} markers indicate the landmarks. Points $x \in X_e$ that do not contribute to the flooding of the edge $e = \{l_1,l_2\}$ are grayed out.
    \textbf{Left}: All simplices have distinct filtration values. Thus, at the shown radius, there are two connected components, $(\{l_1\}$ and $\{l_2\})$.
    \textbf{Right}: $\{l_2\}$ and $\{l_1,l_2\}$ have equal filtration value, and $\forall r\ge 0$, there is at most one connected component.}
    \label{fig:masked_flood_edge}
    \vspace{-0.2cm}
\end{figure}

\end{document}